\documentclass[journal]{IEEEtran}
\usepackage{amsmath,amsfonts}
\usepackage{algorithmic}
\usepackage{array}
\usepackage[caption=false,font=normalsize,labelfont=sf,textfont=sf]{subfig}
\usepackage{textcomp}
\usepackage{stfloats}
\usepackage{url}
\usepackage{verbatim}
\usepackage{graphicx}
\def\BibTeX{{\rm B\kern-.05em{\sc i\kern-.025em b}\kern-.08em
    T\kern-.1667em\lower.7ex\hbox{E}\kern-.125emX}}
\usepackage{balance}
\usepackage{booktabs}
\usepackage{multirow}
\usepackage{algorithm}
\usepackage{xcolor}
\usepackage{amsmath}
\usepackage{bbding}
\usepackage{pifont}
\newcommand{\cmark}{\ding{51}}
\newcommand{\xmark}{\ding{55}}
\newcommand{\OUTPUT}{\item[\textbf{Output:}]}
\usepackage[normalem]{ulem}

\begin{document}
\title{SCALMU: Synthetically-trained Coupling of Adaptive Learned Multiplicative Updates for Hyperspectral-Multispectral Fusion}

\author{Xinxin Xu, Yann Gousseau, Christophe Kervazo, Saïd Ladjal
\thanks{The authors are with LTCI, Télécom Paris, Institut Polytechnique de Paris, 91120 Palaiseau, France (e-mail: xinxin.xu@telecom-paris.fr; yann.gousseau@telecom-paris.fr; christophe.kervazo@telecom-paris.fr; said.ladjal@telecom-paris.fr).}
}

\markboth{}
{SCALMU}
\maketitle
\begin{abstract}

HyperSpectral-MultiSpectral Image (HSI-MSI) fusion aims to recover a high-resolution hyperspectral image from a low-resolution HSI and a high-resolution MSI. Classical methods such as Coupled Nonnegative Matrix Factorization (CNMF) benefit from a strong physical interpretability but suffer from inferior results compared to their deep-learning counterparts. To address this limitation, we propose SCALMU (Synthetically-trained Coupling of Adaptive Learned Multiplicative Updates), a novel blind unrolled neural network architecture that integrates adaptive learnable matrices within the classical framework of CNMF multiplicative updates, improving its results. Due to its architectural proximity with CNMF, the resulting algorithm preserves physical interpretability and nonnegativity constraints. To overcome the scarcity of supervised training data, we generate a synthetic HSI-MSI dataset using the dead leaves model and train SCALMU end-to-end under synthetic supervision. Experiments on several datasets show that SCALMU outperforms state-of-the-art methods and highlights the potential of blind fusion trained with synthetic data. The code is available at
\url{https://github.com/xinxinxu99/SCALMU.git}

\end{abstract}


\section{Introduction}
\label{sec_intro}
\IEEEPARstart {H}yperSpectral (HSI) and MultiSpectral (MSI) images are three-dimensional data cubes with two spatial and one spectral dimensions. When both modalities are used to acquire the same scene, the resulting images differ in spectral richness and spatial detail. HSIs contain hundreds of contiguous narrow bands, enabling precise material identification and supporting diverse applications such as source separation \cite{fahes2022unrolling}, target detection \cite{wu2022uiu}, vegetation monitoring \cite{mills2010evaluation}, and land cover classification \cite{yao2023extended}. However, due to physical constraints, there is a tradeoff between spatial and spectral resolutions, leading in HSIs  to a low spatial resolution, which is further amplified in remote sensing due to large sensor-to-scene distances. In contrast, MSIs capture only a few broad spectral bands but at much higher spatial resolution, offering finer spatial structures and sharper details.

Combining these complementary modalities, namely the high spectral precision of HSI and the high spatial details of MSI, defines the HSI–MSI fusion problem, also referred to as hyperspectral image super-resolution. The goal is to reconstruct a high-resolution hyperspectral image that preserves both the spectral fidelity of HSI and the spatial richness of MSI, a capability that has become increasingly important for modern remote sensing and environmental applications \cite{lu2020recent}.

Existing HSI–MSI fusion methods can be broadly categorized into model-based and deep learning approaches \cite{yokoya2017hyperspectral}. Traditional model-based techniques, exemplified by the coupled nonnegative matrix factorization (CNMF) \cite{yokoya2011coupled}, decompose the HSI into high spectral-resolution endmembers and leverage MSI-derived high spatial-resolution abundances through a linear unmixing model. This physically interpretable approach ensures spectral fidelity without requiring ground-truth HR-HSI data. However, CNMF requires the prior knowledge of degradation operators and suffers from slow convergence through iterative multiplicative updates.

More recently, the advent of deep learning has enabled fully unsupervised and blind fusion methods that jointly estimate these degradation operators from data. Architectures like HyCoNet \cite{zheng2020coupled} and MIAE \cite{liu2022model} integrate physical modeling within neural networks, learning degradation parameters alongside the fusion process to enhance the reconstruction quality and flexibility. 

Despite these advances, deep learning approaches depend on scarce real training pairs with limited generalization. In this work, we propose to overcome this limitation by relying on synthetic training sets. For the super-resolution of RGB images, Achddou et al. \cite{achddou2023fully} demonstrated the feasibility of training with purely synthetic images. Recently, we proposed a dead leaves model based dataset for single hyperspectral image super-resolution \cite{xu2026synthetic}, establishing the potential of synthetic training for hyperspectral restoration tasks. 

In this paper,  in the framework of HSI-MSI fusion tasks, we introduce SCALMU (Synthetically-trained Coupling of Adaptive Learned Multiplicative Updates), which unrolls CNMF multiplicative updates into a learnable architecture trained exclusively on dead leaves synthetic data, combining physical interpretability with accurate data-driven reconstruction. The main contributions are as follows:

\begin{itemize}
    \item We propose CALMU, an unrolled CNMF algorithm that integrates learnable components within the classical framework of multiplicative updates. This unrolled design preserves the interpretability of CNMF while requiring far fewer iterations and improving the super-resolution results.
    
    \item We introduce a synthetic training set generation pipeline for hyperspectral–multispectral fusion based on the dead leaves model and hyperspectral unmixing, allowing the creation of realistic spatial and spectral patterns with full ground-truth control. By training on synthetic data, our method enables fully end-to-end learning without requiring ground-truth high-resolution hyperspectral images.
    
    \item The main novelty of SCALMU lies in the combination of three complementary ingredients within a unified framework: adaptive learned multiplicative updates (CALMU), synthetic training based on dead leaves generation, and blind estimation of the spatial and spectral degradation operators. Rather than introducing these components independently, our contribution is to show that their coupling yields an interpretable, trainable, and effective approach for HSI--MSI fusion.
    
    \item We demonstrate that combining our physics-informed unrolled architecture (CALMU) with a purely synthetic training strategy yields superior super-resolution performance compared to state-of-the-art model-based and deep learning methods. Extensive experiments on multiple datasets validate that our approach generalizes exceptionally well to real-world scenarios, offering a highly efficient and practical solution for HSI-MSI fusion.
    
\end{itemize}

The rest of this paper is organized as follows. In Section \ref{sec_related_work}, we give an overview of the related work. Section \ref{proposed_method} details the proposed SCALMU framework, presenting its two main components: CALMU, the unrolled CNMF formulation, and the procedure for the Synthetic dataset generation leading from CALMU to SCALMU. Section \ref{Experiments} reports experimental results, comparisons with state-of-the-art methods, and ablation studies, demonstrating the effectiveness of our approach.
Finally, conclusions and perspectives are drawn in Section \ref{conclusion}.

\section{Related Work}
\label{sec_related_work}

HSI-MSI fusion has been extensively studied over the past decades, evolving from physically-grounded model-based methods to data-driven deep learning approaches. This section reviews model-based fusion techniques (Section \ref{subsec_model_based}), deep learning-based methods, addressing supervised and unsupervised paradigms (Section \ref{subsec_DL_based}), and closely related to our work, recent advances in synthetic data generation to overcome training data scarcity (Section \ref{subsec_synthetic}).

\subsection{Model based fusion method}
\label{subsec_model_based}

Model-based methods for HSI–MSI fusion explicitly make use of physical observation models and handcrafted priors on the latent high-resolution hyperspectral image. Early approaches benefited from spectral unmixing and Maximum A Posteriori estimation with stochastic mixing models. Pioneering works include Zhukov et al. \cite{zhukov1999unmixing}, who proposed multisensor multiresolution fusion by linearly unmixing HSI into endmembers and using MSI high-resolution abundances ; Eismann et al. \cite{eismann2004application}, who applied stochastic mixing models assuming pixel-wise endmember variability for HSI resolution enhancement; and Hardie et al. \cite{hardie2004map}, who developed MAP estimators fusing HSI with auxiliary high-resolution MSI via joint spatial-spectral regularization. These methods were extended to arbitrary spectral responses \cite{eismann2005hyperspectral}.

A major characteristic of these early model-based approaches was to represent hyperspectral data in a low-dimensional subspace. Under this assumption, fusion can be formulated through matrix or tensor factorization, dictionary learning, or spectral unmixing into endmembers and abundances. For instance, dictionary-based methods decompose HSI into spectral atoms and sparse coefficients \cite{kawakami2011high}, while Akhtar et al.~\cite{akhtar2014sparse} learned overcomplete dictionaries by incorporating sparsity, nonnegativity, and spatial priors through greedy pursuit. In these formulations, fusion amounts to jointly estimating spectral bases and spatial coefficients so that the reconstructed image, after spatial and spectral degradation, matches the observed HSI and MSI. Recent extensions also include noise-aware and robust matrix-factorization formulations, such as MixFus \cite{fu2023mixed}, which explicitly models mixed noise during HSI-MSI fusion, and GSFus \cite{fu2021fusion}, which introduces group sparsity and a plug-and-play denoising prior to handle localized discrepancies while improving robustness to noise.

Among these, CNMF \cite{yokoya2017hyperspectral, yokoya2011coupled} stands out by decomposing each pixel into high spectral-resolution endmembers from HSI and high spatial-resolution abundances from MSI, enforcing nonnegativity constraints and coupling both modalities through degradation operators to ensure spectral-spatial consistency without requiring ground-truth HR-HSI. Despite their robustness and interpretability, these methods often require iterative optimization that can be computationally expensive and sensitive to initialization or degradation model inaccuracies.

\subsection{Deep-learning based fusion method}
\label{subsec_DL_based}

Deep-learning-based methods for HSI–MSI fusion have recently emerged as powerful alternatives to model-based frameworks, replacing handcrafted physical priors with data-driven representations learned end-to-end from paired HSI–MSI data \cite{wang2017deep}. Instead of explicitly modeling spectral mixing or imposing spatial regularization, deep networks directly learn nonlinear mappings from low-resolution HSI and high-resolution MSI to a high-resolution HSI, effectively capturing complex spatial–spectral correlations \cite{xie2019multispectral, zheng2020coupled}. This paradigm shift from analytical optimization to learned inference has led to notable improvements in reconstruction accuracy and robustness to unknown degradations \cite{zhang2020unsupervised}, though it also introduces a strong dependence on the fidelity of training pairs and the realism of the simulated degradations.

Early supervised deep learning approaches primarily relied on residual convolutional neural networks (CNNs). Wang et al. introduced a deep residual CNN that jointly processes HSI and MSI inputs, predicting high-frequency spatial details through skip connections while preserving global spectral structure \cite{wang2017deep}. This residual learning paradigm inspired subsequent architectures like Han et al.'s SSF-CNN, which fuses spatial high-resolution and spectral low-resolution features within shared residual blocks \cite{han2018ssf}. In the same vein, Dian et al. \cite{dian2020regularizing} regularized HSI-MSI fusion with a CNN denoiser prior, improving reconstruction quality while enhancing robustness to noise. Later developments emphasized multi-scale feature extraction and high-resolution guidance: Xie et al.'s MS/HS FusionNet  \cite{xie2019multispectral} combines unrolled model-based priors with residual fusion modules and progressive reconstruction, while Ran et al.'s GuidedNet  \cite{ran2023guidednet} leverages a high-resolution guidance branch alongside feature reconstruction for enhanced spatial–spectral fidelity.

Unsupervised/self-supervised methods emerged to address the need for paired training data, which comprehend the HSI, the MSI and the corresponding high-reolution ground-truth HSI, incorporating physical modeling directly into neural architectures. Qu et al.'s uSDN employs coupled encoder-decoders with sparse Dirichlet priors \cite{qu2018unsupervised}, while HyCoNet \cite{zheng2020coupled} adopts a three-branch autoencoder capable of jointly estimating spatial and spectral degradations from observations. Similarly, MIAE \cite{liu2022model} introduces model-inspired autoencoders, UDALN \cite{li2022deep} explores blind degradation learning within an unsupervised deep fusion framework, BUSIFusion \cite{li2023busifusion} proposes a blind unsupervised single-image fusion strategy for hyperspectral and RGB observations, and EU2ADL \cite{gao2023enhanced} enhances degradation learning with attention mechanisms. CSSnet \cite{li2023hyperspectral} leverages cross-scale nonlocal attention, and DCnet \cite{hong2023decoupled} proposes decoupled fusion separating common and sensor-specific components. More recently, advanced unsupervised methods have pushed the boundaries further. Zhang et al.'s test-time adaptation \cite{zhang2024unsupervised} enables handling unknown degradations through self-supervised fine-tuning at inference time. Symmetrical propagation \cite{li2022symmetrical} captures long-range spatial-spectral dependencies via bidirectional feature propagation. Transformer-based FusFormer \cite{hu2022fusformer} leverages global self-attention for enhanced cross-modality interactions, while FeafusFormer \cite{cao2024unsupervised} further investigates blind fusion through complementary local and global feature modeling. HSR-Diff \cite{wu2023hsr} introduces conditional diffusion models for iterative high-fidelity reconstruction. EDIP-Net \cite{li2025enhanced} improves deep image priors with zero-shot input generation and dual U-Net architectures. Finally, OTIAS \cite{deng2025otias} proposes OcTree-based adaptive sampling to preserve fine details across scales.

\subsection{Synthetic data}
\label{subsec_synthetic}

Synthetic datasets are particularly relevant for HSI-MSI fusion as they provide paired training data with perfect ground-truth control, overcoming the scarcity of real high-resolution HSI references. However, their use remains uncommon in the hyperspectral imaging and remote sensing literature. In contrast, in the field of computer vision, the use of synthetic data has been commonplace, and many works have been devoted to this topic, mostly for classification, estimation or detection tasks, relying for instance on simplified object generation \cite{dosovitskiy2015flownet} or more recently text-to-image models \cite{tian2023stablerep}. For image restoration tasks, various generative models have been proposed to simulate realistic image statistics and structures. Markov random fields \cite{cross1983markov} allow the modeling of local spatial dependencies, making them effective for capturing texture patterns. Wavelet-based models \cite{heeger1995pyramid} are well-suited for representing images at multiple scales, enabling the synthesis of textures with both fine and coarse structures. Gaussian models \cite{galerne2011micro} offer a mathematically tractable way to reproduce second-order statistics of natural scenes, which often suffice for basic texture synthesis. The dead leaves model offer a simple framework for simulating edges and homogeneous regions by overlapping opaque objects and effectively reproduce non-Gaussian statistics observed in natural images \cite{matheron1968modele, alvarez1999size, gousseau2003dead, bordenave2006dead}. Achddou et al.  demonstrated that convolutional neural networks trained exclusively on dead leaves images can achieve competitive performance for the denoising and  super-resolution of natural images \cite{achddou2023fully}. Recently, we extended this approach to single hyperspectral image restoration in \cite{xu2026synthetic}, introducing a dead leaves synthetic dataset with abundance map generation that enables unsupervised single image super-resolution training.

\section{Proposed method}
\label{proposed_method}

This section presents the SCALMU framework for blind HSI-MSI fusion. Subsection \ref{subsec_formulation} formulates the problem, \ref{subsec_network_structure} presents CALMU (the unrolled CNMF network), \ref{subsec_dataset_generation} describes our synthetic data generation leading to SCALMU, and \ref{subsec_blindnet} details blind degradation estimation.

\begin{figure*}
    \centering
    \includegraphics[width=1\linewidth]{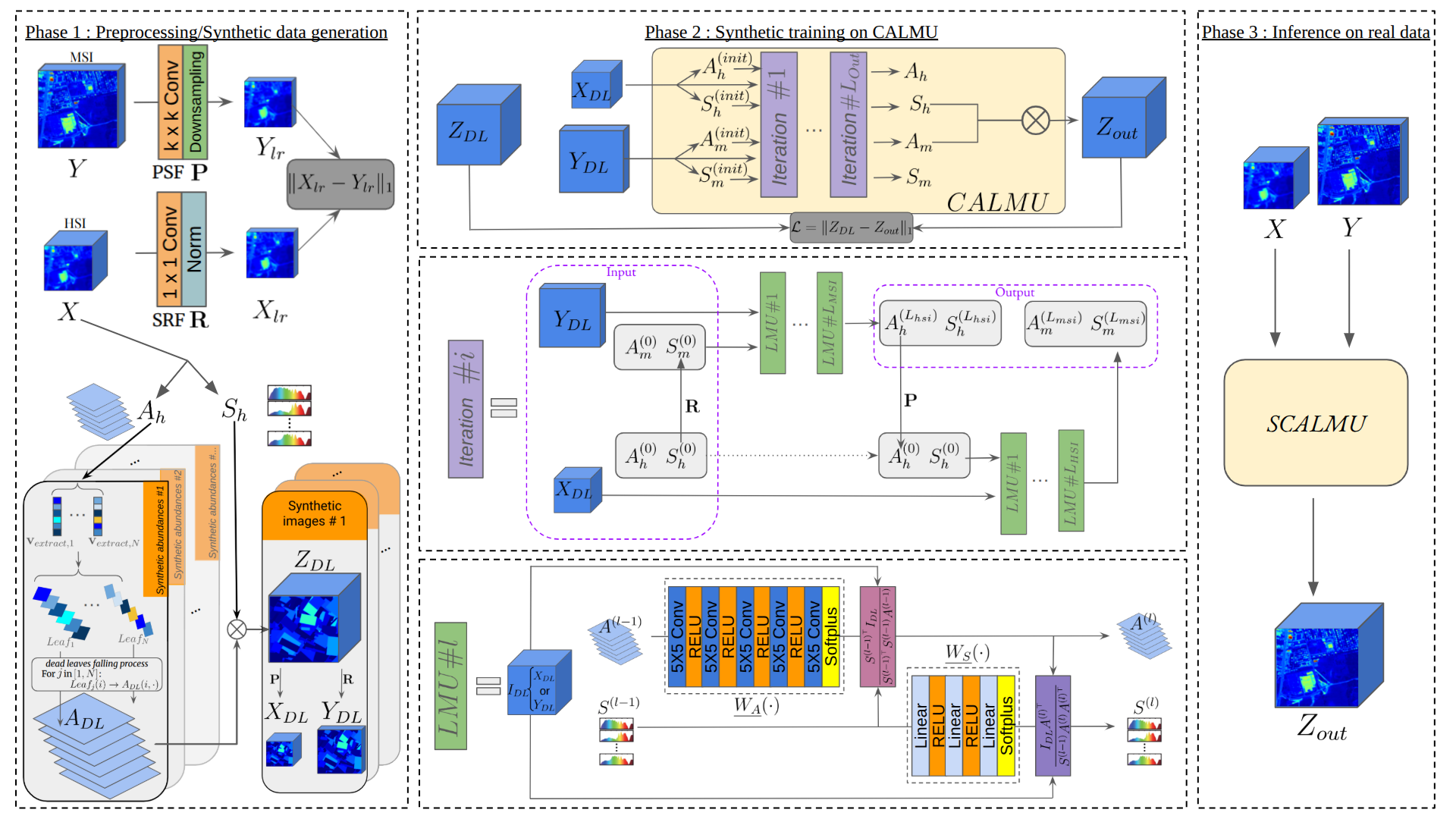}
    \caption{Structure of the proposed SCALMU: Synthetically-trained Coupling of Adaptive Learned Multiplicative Updates~: Phase 1 consists in the blind determination of degradations $\mathbf{P}$ and $\mathbf{R}$ followed by the dead leaves dataset generation. Phase 2 (central part) is the Training using the dead leaves dataset. Phase 3 corresponds to test time~: merely replace $X_{DL}$ and $Y_{DL}$ by real hyper and multi-spectral images in phase 2, $Z$ (high resolution hyperspectral image) being the output of the network.} 
    \label{fig_SCALMU_framework}
\end{figure*}

\subsection{Problem Formulation}
\label{subsec_formulation}

HSI-MSI fusion aims to estimate a super-resolved image, denoted as $ Z \in \mathbb{R}^{B \times HW} $, from the same scene observed by an HSI, $ X \in \mathbb{R}^{B \times hw} $, and a MSI, $ Y \in \mathbb{R}^{b \times HW} $. Here, $(h, w)$ and $(H, W)$ represent the spatial dimensions of the HSI and MSI, respectively, with $h < H$ and $w < W$, while $B$ and $b$ denote the number of spectral bands in the HSI and MSI, satisfying $b < B$. By introducing the point spread function (PSF)\footnote{$\mathbf{P}$ is, in fact, the linear operation consisting in blurring an image with a PSF and then downsampling it. The linear operation notation is more convenient for our presentation and details about the determination of the actual PSF are given in section \ref{subsec_blindnet} } $\mathbf{P} \in \mathbb{R}^{hw \times HW}$ and the spectral response function (SRF) $\mathbf{R} \in \mathbb{R}^{b \times B}$, which model the spatial and spectral degradations of $Z$, respectively, the HSI $X$ and the MSI $Y$ can be expressed as:  
\begin{equation}
\label{eq_PR}
\begin{aligned}
X &= Z \mathbf{P}^\top , \\
Y &= \mathbf{R} Z ,
\end{aligned}
\end{equation}

In parallel, under the linear hyperspectral unmixing model \cite{yokoya2011coupled}, by introducing $M$ as the number of materials, the observations $X$ and $Y$ can be expressed as $X \simeq S_h A_h$ and $Y \simeq S_m A_m$, where $S_h \in \mathbb{R}^{B \times M}$ and $S_m \in \mathbb{R}^{b \times M}$ denote the endmember matrices obtained from the HSI and MSI, respectively, while $A_h \in \mathbb{R}^{M \times hw}$ and $A_m \in \mathbb{R}^{M \times HW}$ denote the abundance maps obtained from the HSI and MSI, respectively. The fused image can then be obtained as $Z = S_h A_m$, where $S_h$ and $A_m$ encode the high spectral resolution (from the HSI) and high spatial resolution (from the MSI) information, respectively. Based on this model and Eq. \eqref{eq_PR}, we obtain:

\begin{equation}
\label{eq_unmixing_fusion}
\begin{aligned}
X & \simeq S_h A_h = S_h A_m \mathbf{P}^\top, \\
Y & \simeq S_m A_m = \mathbf{R} S_h A_m, \\
Z &= S_h A_m ,
\end{aligned}
\end{equation}
with the coupling relations $A_h = A_m \mathbf{P}^\top$ and $S_m = \mathbf{R} S_h$.

\subsection{Proposed CALMU Structure}
\label{subsec_network_structure}

We propose CALMU, an unrolled version of the CNMF algorithm \cite{yokoya2011coupled}. We recall that in this method the fusion problem is formulated as a joint minimization, under nonnegativity conditions, of the following cost functions:  
\begin{equation}
\label{eq_cost_fusion}
\begin{aligned}
\underset{S_h,A_h\geq 0}{\mathrm{argmin}} & \|X - S_h A_h\|_F^2 ,\\
\underset{S_m,A_m \geq 0}{\mathrm{argmin}} & \|Y - S_m A_m\|_F^2 ,
\end{aligned}
\end{equation}

where $\| \cdot\|_F$ denotes the Frobenius norm. In the CNMF algorithm \cite{yokoya2011coupled}, the minimization is done iteratively using the well-known Multiplicative Updates (MU) \cite{lee1999learning} and leveraging the coupling constraints $A_h = A_m \mathbf{P}^\top$ and $S_m = \mathbf{R} S_h$.
In the present work, the MU iterations of CNMF are unrolled building on the Recursive Adaptive Learned Multiplicative Updates (RALMU) framework of Kervazo et al. \cite{kervazo2026Unrolled} initially designed for hyperspectral unmixing.

More precisely, we introduce four lightweight neural networks in CNMF updates to predict some adaptive learnable matrices $ \underline{W_{A_m}}(\cdot), \underline{W_{S_m}}(\cdot), \underline{W_{A_h}}(\cdot)$ and $ \underline{W_{S_h}}(\cdot) $, which dynamically adjust the MU update rules based on the input data characteristics, enabling to largely improve CNMF super-resolution results while dramatically reducing the number of required iterations. As illustrated in the $LMU$ block of Fig. \ref{fig_SCALMU_framework},  the estimation of the abundance-related matrices is handled by two convolutional networks, one dedicated to the multispectral branch $ \underline{W_{A_m}}(\cdot) $ and the other to the hyperspectral branch $ \underline{W_{A_h}}(\cdot) $. Each of these networks is implemented as a 5-layer Conv2D architecture with $5 \times 5$ convolutions, intermediate ReLU activations, and a final Softplus activation to enforce nonnegativity. They process the current abundance estimates to generate spatially adaptive weights, allowing the model to refine abundance maps at each iteration. The endmember-related matrices are controlled by two MLPs that produce the corresponding spectral transformations, denoted as $\underline{W_{S_m}}(\cdot) $ and $ \underline{W_{S_h}}(\cdot) $, each composed of three linear layers with ReLU activations between hidden layers and a final Softplus activation. These MLPs operate along the spectral dimension and learns to adjust the endmember updates according to the spectral distribution of the input. To maintain the nonnegativity in the unrolled multiplicative updates, all predicted matrices are constrained through the final Softplus layer. Due to the structure of its alternating updates, the architecture ensures that its intermediate outputs maintain their physical meaning as abundance and endmember matrices, contrarily to methods relying on abstract latent spaces.

The entire CALMU network, described in Algorithm \ref{algo_CNMF_unrolling} and illustrated in Phase 2 of Fig. \ref{fig_SCALMU_framework}, is trained end-to-end using synthetic data by employing the L1 reconstruction loss $\mathcal{L} = \|Z - Z_{out}\|_1$, ensuring full control over the ground-truth super-resolved images and the associated degradation processes. The procedure used to generate these synthetic data is detailed in the following subsection.

\begin{algorithm}
\caption{CALMU algorithm}  
\label{algo_CNMF_unrolling}
\begin{algorithmic}[1]
\REQUIRE HSI $X$, MSI $Y$, PSF $\mathbf{P}$, SRF $\mathbf{R}$, number of iterations $L_{MSI}$, $L_{HSI}$, $L_{Out}$  

\STATE \textbf{\textit{Step 1:}} Initial estimation with Multiplicative Update (MU) \\  
$A_h^{(L_{HSI})}, S_h^{(L_{HSI})} \leftarrow MU(X)$ \\
$ A_m^{(L_{MSI})} \leftarrow rand_{[0,1]}(\cdot) $  

\STATE \textbf{\textit{Step 2:}} Multispectral update \\  
$S_m^{(0)} \leftarrow \mathbf{R} S_h^{(L_{HSI})} \quad  A_m^{(0)} \leftarrow  A_m^{(L_{MSI})}$
\FOR{$l = 0$ \textbf{to} $L_{MSI}-1$}  
    \STATE $A_m^{(l+1)} \leftarrow A_m^{(l)} \odot \underline{W_{A_m}}(A_m^{(l)}) \odot \frac{S_m^{(l)^\top} Y}{S_m^{(l)^\top} S_m^{(l)} A_m^{(l)}}$  
    \STATE $S_m^{(l+1)} \leftarrow S_m^{(l)} \odot \underline{W_{S_m}}(S_m^{(l)}) \odot \frac{Y A_m^{(l+1)^\top}}{S_m^{(l)} A_m^{(l+1)} A_m^{(l+1)^\top}}$  
\ENDFOR  

\STATE \textbf{\textit{Step 3:}} Hyperspectral update \\  
$A_h^{(0)} \leftarrow A_m^{(L_{MSI})}  \mathbf{P}^T$, \\ 
$S_h^{(0)} \leftarrow S_h^{(L_{HSI})} \odot \frac{X A_h^{(0)^\top}}{S_h^{(L_{HSI})} A_h^{(0)} A_h^{(0)^\top}}$

\FOR{$l = 0$ \textbf{to} $L_{HSI}-1$}  
    \STATE $A_h^{(l+1)} \leftarrow A_h^{(l)} \odot \underline{W_{A_h}}(A_h^{(l)}) \odot \frac{S_h^{(l)^\top} X}{S_h^{(l)^\top} S_h^{(l)} A_h^{(l)}}$  
    \STATE $S_h^{(l+1)} \leftarrow S_h^{(l)} \odot \underline{W_{S_h}}(S_h^{(l)}) \odot \frac{X A_h^{(l+1)^\top}}{S_h^{(l)} A_h^{(l+1)} A_h^{(l+1)^\top}}$   
\ENDFOR  

\STATE \textbf{\textit{Step 4:}} Repeat Steps 2 and 3 for $L_{Out}$ iterations  

\RETURN $Z_{out} = S_h^{(L_{HSI})} A_m^{(L_{MSI})}$ 
\end{algorithmic} 
\smallskip  
\noindent \textbf{Notes:}\textit{ $\odot$ denotes element-wise multiplication. The case where the four $\underline{W^{(l)}}$ matrices are set to 1 (non-learnable) corresponds to the original CNMF updates.} 
 
\end{algorithm}

\subsection{Synthetic Dataset Generation by the dead leaves model}
\label{subsec_dataset_generation}

To train CALMU, we generate synthetic HSI-MSI pairs $(X_{DL},Y_{DL})$ with synthetic ground-truth $Z_{DL}$ using abondances from the low-resolution HSI $X$ and the dead leaves model. This stochastic model generates images through sequential superimposition of random shapes, referred to as leaves, at random positions \cite{matheron1968modele, bordenave2006dead}, reproducing key statistical properties of natural images \cite{alvarez1999size, lee2001occlusion, gousseau2007modeling}. Leaf shapes follow random geometric models with positions distributed via a stationary Poisson point process over the spatial domain. The process continues until a stationary state is reached, which can be efficiently achieved via perfect simulation techniques \cite{kendall1999perfect}.

Taking inspiration from our previous work \cite{xu2026synthetic}, we develop a methodology to generate pairs of synthetic HSI and MSI from the low-resolution hyperspectral image $X$ only. Here again, our approach builds upon a first hyperspectral unmixing preprocessing step (performed through the original MU), during which we estimate the endmembers and abundances $S_h, A_h = MU(X)$ from the HSI.

\begin{figure}
    \centering
    \includegraphics[width=1\linewidth]{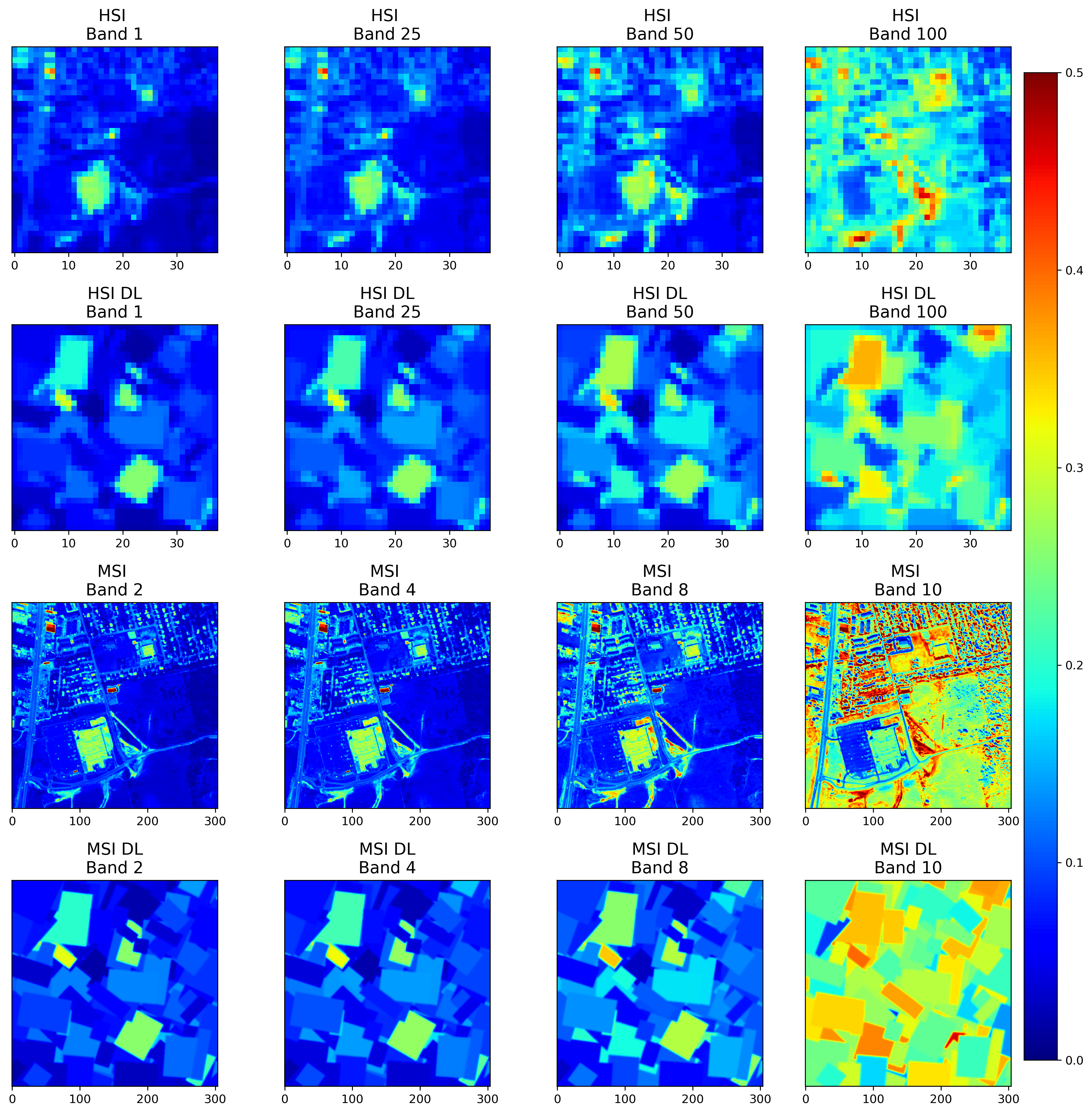}
    \caption{Visual comparison between real and synthetic Urban dataset observations. Top two rows: HSI $X$ and $X_{DL}$ (synthetic) at spectral bands 1, 25, 50, and 100. Bottom two rows: MSI $Y$ and $Y_{DL}$ (synthetic) at bands 2, 4, 8, and 10. The real observations $X$ and $Y$ were simulated from the original-resolution HSI, taken as the reference image $Z$ at its native spatial resolution, using an $\times 8$ Gaussian PSF and Sentinel-2 SRF.} 
    \label{fig_DL}
\end{figure}

Starting from these low-resolution abundances, we construct the high-resolution synthetic abundances $A_{DL}$ by superimposing rectangular leaf-shaped regions according to the dead leaves process. Following the same parameterization as in our previous work \cite{xu2026synthetic}, each leaf is characterized by its width and height $(a,b)$, orientation $\theta$, center position $(x,y)$, and an abundance vector randomly selected from $A_h$. The rectangle dimensions $a$ and $b$ are independently sampled from a uniform distribution over $[2r,\min(H,W)/3]$, where $r$ denotes the super-resolution factor, while $\theta$ is uniformly drawn in $[0^\circ,45^\circ]$ and the leaf centers positions $(x,y)$ are uniformly sampled over the spatial domain. Leaves are sequentially superimposed until all pixels have been covered at least once. Each leaf inherits its abundance values from randomly selected vectors of $A_h$, ensuring consistency with the physical abundance proportions and preserving the nonnegativity and sum-to-one constraints.  
By reusing the endmembers matrix $S_h$, we then reconstruct the high-resolution synthetic scene as $Z_{DL} = S_h A_{DL}$. The corresponding degraded hyperspectral and multispectral observations are subsequently obtained through the application of the spatial and spectral degradation operators: $X_{DL} = Z_{DL} \mathbf{P}^\top$ and $Y_{DL} = \mathbf{R} Z_{DL}$ which are automatically estimated (cf. Section \ref{subsec_blindnet}). Phase 1 of Fig. \ref{fig_SCALMU_framework} shows the integration of the synthetic dataset creation into SCALMU's preprocessing step. An overview of this synthetic data generation is illustrated in Fig. \ref{fig_DL}, visually comparing the real observations $X$ and $Y$ with their synthetic counterparts $X_{DL}$ and $Y_{DL}$. The detailed procedure is provided in Algorithm \ref{algo_dead_leaves}.

The use of an unmixing step has significant advantages compared to the direct generation of synthetic high-resolution HSIs. First, it reduces computational resources by making the dead leaves generation operate in a low-dimensional abundance space rather than using the full high-dimensional spectra. Second, it ensures sensor independence by decoupling spatial patterns from sensor-specific properties like varying band counts or spectral responses. Finally, it provides a physically interpretable decompositions that preserve nonnegativity and sum-to-one constraints over the abundances, enhancing realism and consistency.

\begin{algorithm} 
\caption{Synthetic dataset Generation with the dead leaves model}
\label{algo_dead_leaves}
\begin{algorithmic}[1]
\REQUIRE Low-resolution HSI $X \in \mathbb{R}^{B \times hw}$, PSF $\mathbf{P}$, SRF $\mathbf{R}$, target resolution $(H, W)$, number of endmembers $M$  
\OUTPUT Synthetic high resolution image $Z_{DL}$, synthetic HSI $X_{DL}$, synthetic MSI $Y_{DL}$  

\STATE Estimate endmembers and abundances $(S_h, A_h) \gets MU(X)$
\STATE Initialize synthetic high-resolution abundances $A_{DL} \gets 0 \in \mathbb{R}^{M \times H \times W}$
\STATE Initialize $Mask \gets \emptyset$
\WHILE{$\# Mask < HW$}
    \STATE Randomly draw $a,b$ (rectangle width and height)
    \STATE Randomly draw $\theta$ (orientation angle)
    \STATE Randomly draw $(x,y)$ (center position)
    \STATE Sample $Leaf \gets \text{Rect}(a, b, \theta) + (x,y)$  
    \STATE Extract a material proportion vector $\textbf{v}_{extract} \gets \text{RandomPixel}(A_h)\in\mathbb{R}^M$ 
    \FOR{$(i,j) \in Leaf \backslash Mask$}
        \STATE $A_{DL}[.,i,j] \gets \textbf{v}_{extract}[.]$
    \ENDFOR
    \STATE $Mask \gets Mask \cup Leaf$
\ENDWHILE
\STATE Flatten $A_{DL} \in \mathbb{R}^{M \times H \times W}$ to $A_{DL} \in \mathbb{R}^{M \times HW}$ 
\STATE Compute synthetic $Z_{DL} = S_h A_{DL}$
\STATE Compute synthetic HSI $X_{DL} = Z_{DL} \mathbf{P}^\top$
\STATE Compute synthetic MSI $Y_{DL} = \mathbf{R} Z_{DL}$
\RETURN $X_{DL}, Y_{DL}, Z_{DL}$
\end{algorithmic}
\end{algorithm}

\subsection{Blind Estimation Network}
\label{subsec_blindnet}

To enable the end-to-end blind use of SCALMU without the prior knowledge of the degradation operators $\mathbf{P}$ and $\mathbf{R}$, we use a blind estimation network that jointly estimates the PSF and SRF from the low-resolution HSI $X$ and high-resolution MSI $Y$, following the unsupervised degradation estimation principles from recent works such as HyCoNet \cite{zheng2020coupled}, MIAE \cite{liu2022model}, and EDIP-Net \cite{li2025enhanced}, as illustrated in Phase 1 of Fig. \ref{fig_SCALMU_framework}. The network parameterizes learnable SRF and PSF kernels via convolution layers without bias, optimizing an L1 loss $\mathcal{L} = \|X_{lr} - Y_{lr}\|_1$ between the spectrally degraded HSI $X_{lr}$ and the spatially blurred-downsampled MSI $Y_{lr}$, while normalization enforces nonnegativity and sum-to-one constraints. The resulting estimates $\mathbf{R}$ and $\mathbf{P}$ are then fed to CALMU and to the synthetic dataset generation process.

\section{Experiments}
\label{Experiments}

This section presents a comprehensive evaluation of SCALMU on standard hyperspectral benchmarks and real satellite data. We first detail the used datasets and the experimental setup (Section \ref{subsec_dataset}). We then compare our approach to  State-of-the-Art (SOTA) methods (Section \ref{subsec_SOTA}), demonstrate real-data generalization (Section \ref{subsec_real_data}), conduct thorough ablation studies on degradation estimation, adaptability, module contributions and relevance of synthetic data (Section \ref{subsec_Ablation}).We also present a preliminary result highlighting the potential of the proposed approach to generalize across synthetic datasets (Section \ref{subsec_generalisation}), before analyzing its computational efficiency (Section \ref{subsec_Efficiency}).

\subsection{Dataset \& Setup}
\label{subsec_dataset}

\begin{figure*}
    \centering
    \includegraphics[width=1\linewidth]{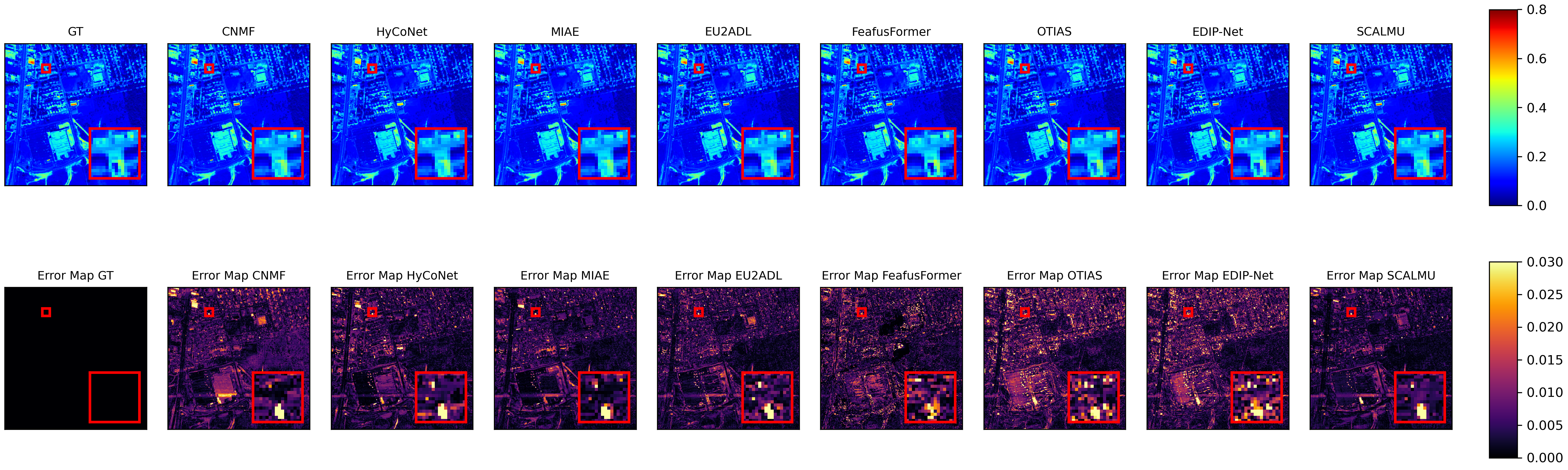}
    \caption{Visual comparison on \textbf{Urban} dataset (band 50). \textbf{Line 1}: reconstructed hyperspectral images (GT, CNMF, HyCoNet, MIAE, EU2ADL, FeafusFormer, OTIAS, EDIP-Net, SCALMU). \textbf{Line 2}: corresponding absolute error maps $|SR - GT|$. Red rectangles indicate regions zoomed in the insets.} 
    \label{fig_SOTA_comparison_Urban}
\end{figure*}

\begin{figure*}
    \centering
    \includegraphics[width=1\linewidth]{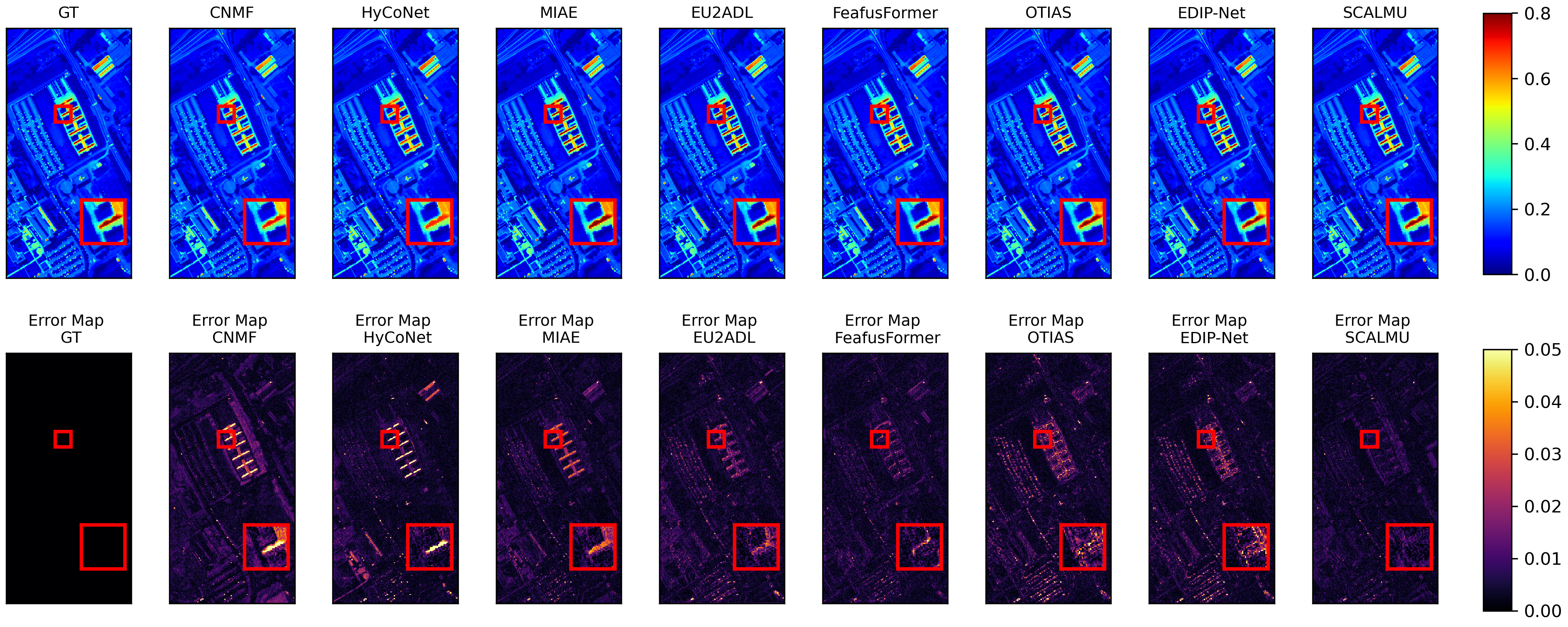}
    \caption{Visual comparison on \textbf{PaviaU} dataset (band 20). \textbf{Line 1}: reconstructed hyperspectral images (GT, CNMF, HyCoNet, MIAE, EU2ADL, FeafusFormer, OTIAS, EDIP-Net, SCALMU). \textbf{Line 2}: corresponding absolute error maps $|SR - GT|$. Red rectangles indicate regions zoomed in the insets.}
    \label{fig_SOTA_comparison_PaviaU}
\end{figure*}

We evaluate our proposed method on four hyperspectral datasets:
\begin{itemize} 
    \item \textbf{Urban}: This dataset was captured by the Hyperspectral Digital Image Collection Experiment (HYDICE) sensor over an urban area near Fort Hood, Texas, USA. It consists of $307 \times 307$ pixels with 210 spectral bands spanning 400–2500 nm at a spectral resolution of 10 nm. After discarding bands affected by water-vapor absorption and low signal quality, 162 bands are retained for evaluation.
    
    \item \textbf{Pavia University (PaviaU)}: Acquired by the ROSIS-3 airborne optical sensor in 2003, this dataset contains $610 \times 340$ pixels with a ground sampling distance of 1.3 m. It covers the 430–840 nm spectral range across 115 bands. After removing 12 bands affected by noise and water-vapor absorption, a $512 \times 256$ subregion corresponding to an urban area and containing the remaining 103 bands is used in our experiments.
    
    \item \textbf{Chikusei}: This dataset was acquired by the VNIR-C hyperspectral sensor over Chikusei, Japan. The original image consists of $2517 \times 2335$ pixels with 128 spectral bands covering the 363–1018 nm wavelength range. For our experiments, a $320 \times 320$ pixel subregion is randomly selected to represent a mixture of agricultural and urban areas.
    
    \item \textbf{PRISMA-Paris}: The PRISMA dataset originates from the \textit{Precursore Iperspettrale della Missione Operativa} (PRISMA) mission, launched by the Italian Space Agency in 2019 \cite{cogliati2021prisma}. Focusing on urban scenes, we use a hyperspectral image acquired over Paris, comprising $256 \times 256$ pixels and 230 spectral bands spanning the 400–2500 nm range. 
\end{itemize}

In all experiments, we adopt a spatial scale factor of $\times 8$ between the MSI and HSI. The original HSI at its native resolution serves as the reference super-resolved image. The low-resolution HSI $X$ is simulated following Wald's protocol \cite{wald1997fusion} by convolving the reference with a Gaussian PSF of standard deviation $\sigma = 3.5$ (using a $15 \times 15$ kernel), followed by bicubic downsampling. For the Urban, Chikusei and PRISMA-Paris datasets, the MSI $Y$ is generated using the Sentinel-2A SRF, comprising 12 bands covering 443–2190 nm. For PaviaU, the MSI $Y$ is generated via the Ikonos SRF, which includes 4 bands spanning 350–1035 nm.

Quantitative performance is assessed using standard hyperspectral image quality metrics: Root Mean Square Error (RMSE), Peak Signal-to-Noise Ratio (PSNR), Spectral Angle Mapper (SAM), Erreur Relative Globale Adimensionnelle de Synthèse (ERGAS), and Universal Image Quality Index (UIQI) \cite{yokoya2017hyperspectral}.

\begin{figure}
    \centering
    \includegraphics[width=1\linewidth]{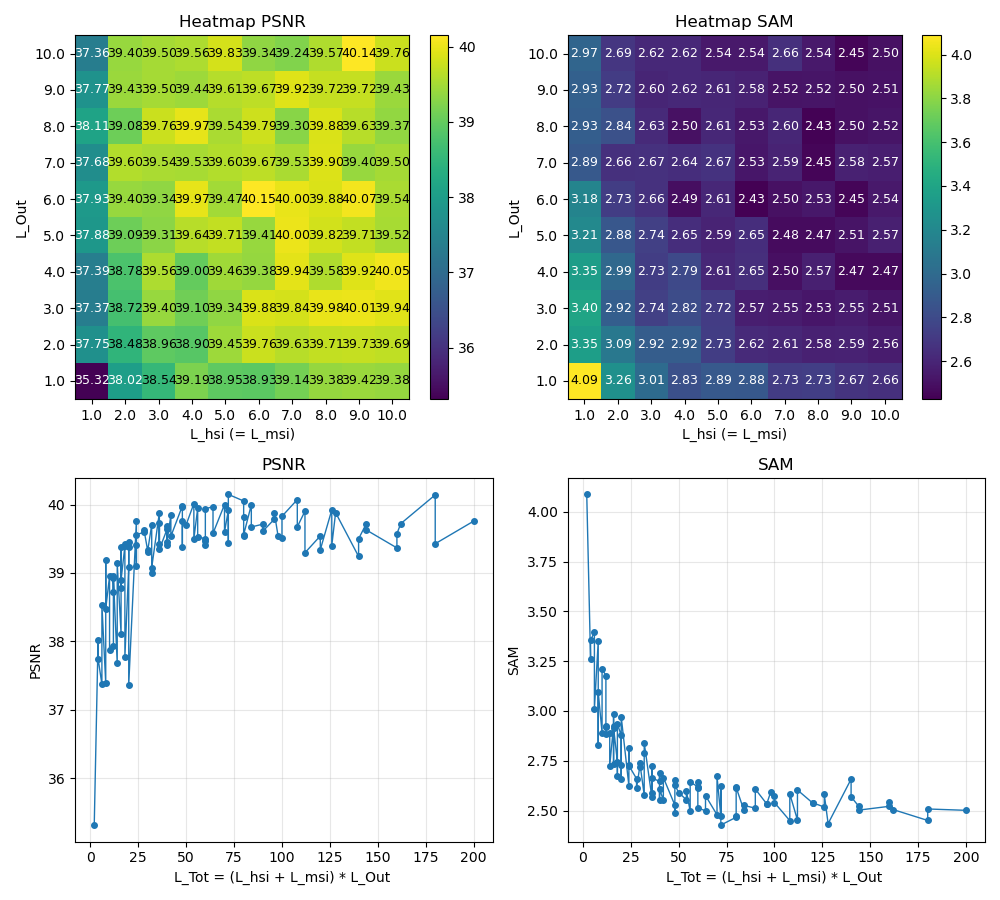}
    \caption{Grid-search analysis of the unrolling depths on the PaviaU dataset. The top row reports heatmaps of PSNR and SAM as functions of $L_{HSI}(=L_{MSI})$ and $L_{Out}$, while the bottom row shows the corresponding evolution with respect to the total number of iterations.}
    \label{fig_grid_search}
\end{figure}

Synthetic training was performed using 1,000 pairs of HSI-MSI image couples $(X_{DL}, Y_{DL})$ generated via the dead leaves model (Section \ref{subsec_dataset_generation}). SCALMU (Algorithm \ref{algo_CNMF_unrolling}) was trained end-to-end using the Adam optimizer with learning rate $\eta = 1 \times 10^{-4}$, employing the L1 reconstruction loss $\mathcal{L} = \|Z_{DL} - Z_{out}\|_1$, where $Z_{out}$, denotes the SCALMU prediction. A grid search was then conducted on the PaviaU dataset over $L_{HSI}, L_{MSI}, L_{Out} \in [1,10]$ to assess the influence of the unrolling depth on reconstruction performance. To limit the computational cost of this exploration, the assumption $L_{HSI} = L_{MSI}$ was adopted. The detailed results are provided in Fig.~\ref{fig_grid_search}, from which the configuration $L_{HSI} = L_{MSI} = L_{Out} = 6$ was retained as the best compromise between convergence speed and reconstruction quality.

\subsection{Comparison with the State-of-the-Art Methods}
\label{subsec_SOTA}

\begin{figure*}
    \centering
    \includegraphics[width=1\linewidth]{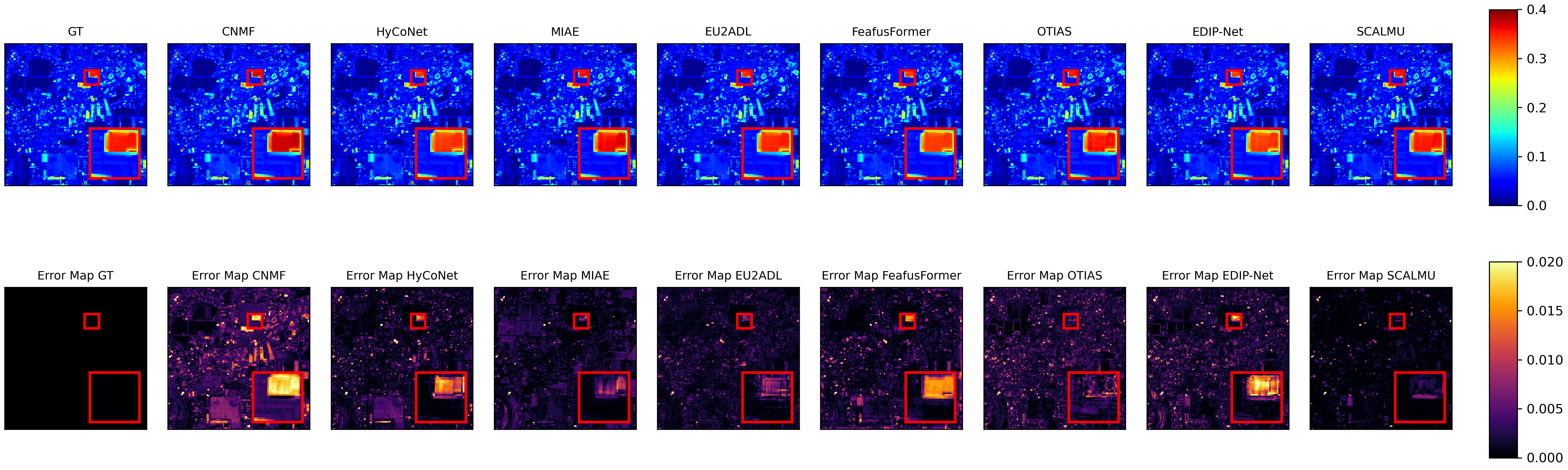}
    \caption{Visual comparison on \textbf{Chikusei} dataset (band 20). \textbf{Line 1}: reconstructed hyperspectral images (GT, CNMF, HyCoNet, MIAE, EU2ADL, FeafusFormer, OTIAS, EDIP-Net, SCALMU). \textbf{Line 2}: corresponding absolute error maps $|SR - GT|$. Red rectangles indicate regions zoomed in the insets.}
    \label{fig_SOTA_comparison_Chikusei}
\end{figure*}

\begin{figure*}
    \centering
    \includegraphics[width=1\linewidth]{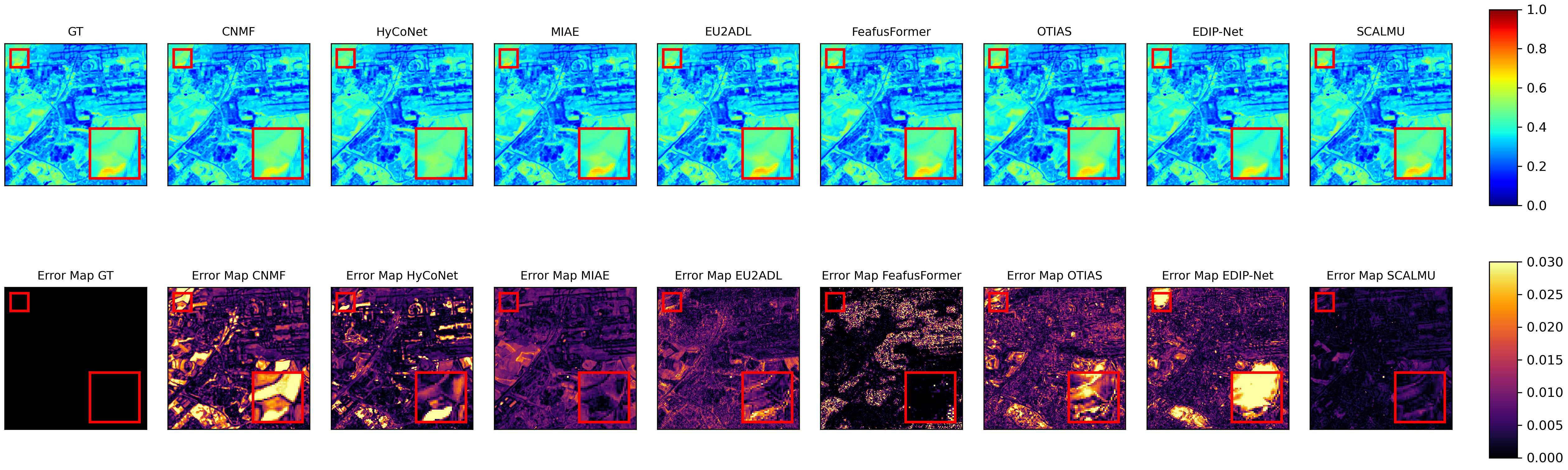}
    \caption{Visual comparison on \textbf{PRISMA-Paris} dataset (band 50). \textbf{Line 1}: reconstructed hyperspectral images (GT, CNMF, HyCoNet, MIAE, EU2ADL, FeafusFormer,  OTIAS, EDIP-Net, SCALMU). \textbf{Line 2}: corresponding absolute error maps $|SR - GT|$. Red rectangles indicate regions zoomed in the insets.}
    \label{fig_SOTA_comparison_Prisma}
\end{figure*}

We compare our method SCALMU, trained exclusively on synthetic data, against state-of-the-art hyperspectral-multispectral fusion methods. Specifically, we evaluate against the classical CNMF baseline \cite{yokoya2011coupled} and six SOTA unsupervised deep learning methods: HyCoNet \cite{zheng2020coupled}, MIAE \cite{liu2022model}, EU2ADL \cite{gao2023enhanced}, FeafusFormer \cite{cao2024unsupervised} ,OTIAS \cite{deng2025otias} and EDIP-Net \cite{li2025enhanced}. All competing methods were tested using the default configurations of the provided codes. While CNMF requires the exact Gaussian PSF ($\sigma=3.5$, $15\times15$ kernel) and corresponding SRF provided during testing, SCALMU and all competing deep learning methods (HyCoNet, MIAE, EU2ADL, FeafusFormer, OTIAS, EDIP-Net) operate in blind configuration without access to degradation parameters, representing their realistic performance in practical scenarios.

Tables \ref{tab_urban_x8}–\ref{tab_prisma_x8} report the quantitative results on the Urban, PaviaU, Chikusei, and PRISMA-Paris datasets at $\times 8$ scale factor. Across all datasets, SCALMU significantly surpasses the classical CNMF baseline, although CNMF uses the non-available in practice ground-truth $\mathbf{P}$ and $\mathbf{S}$. SCALMU also outperforms all its blind deep learning competitors. These results across diverse urban scenes robustly validate our synthetic-training strategy and unrolled CNMF architecture, achieving state-of-the-art performance without requiring degradation model knowledge at test time. Visual comparisons in Figs. \ref{fig_SOTA_comparison_Urban}–\ref{fig_SOTA_comparison_Prisma} confirm SCALMU's advantages through sharper details, better spectral fidelity, and reduced artifacts in error maps.

\begin{table}
\centering
\caption{Quantitative evaluation on \textbf{Urban} dataset ($\times 8$) comparing CNMF, HyCoNet, MIAE, EU2ADL, FeafusFormer, OTIAS, EDIP-Net and SCALMU (ours). Best/second-best results in \textbf{bold}/\underline{underline}.}
\label{tab_urban_x8}
\begin{tabular}{l|ccccc}
\hline
\hline
\textbf{Method} & \textbf{RMSE} $\downarrow$ & \textbf{PSNR} $\uparrow$ & \textbf{SAM} $\downarrow$ & \textbf{ERGAS} $\downarrow$ & \textbf{UIQI} $\uparrow$ \\
\hline
CNMF                & 0.0126&37.99&3.30&1.10&0.9918 \\
HyCoNet               & 0.0101&39.93&2.52&\underline{0.84}&0.9941 \\
MIAE     & 0.0089&41.03&\underline{2.19}&1.21&0.9958 \\
EU2ADL                & \underline{0.0088}&\underline{41.09}&2.29&0.85&\underline{0.9959} \\
FeafusFormer & 0.0096&40.37&2.78&0.85&0.9946\\
OTIAS & 0.0093&40.63&3.00&1.00&0.9944 \\
EDIP-Net & 0.0089&41.03&2.90&1.09&0.9950 \\
SCALMU          & \textbf{0.0084}&\textbf{41.49}&\textbf{2.17}&\textbf{0.82}&\textbf{0.9962}\\
\hline
\hline
\end{tabular}
\end{table}

\begin{table}
\centering
\caption{Quantitative evaluation on \textbf{PaviaU} dataset ($\times 8$) comparing CNMF, HyCoNet, MIAE, EU2ADL, FeafusFormer, OTIAS, EDIP-Net and SCALMU (ours). Best/second-best results in \textbf{bold}/\underline{underline}.}
\label{tab_paviau_x8}
\begin{tabular}{l|ccccc}
\hline
\hline
\textbf{Method} & \textbf{RMSE} $\downarrow$ & \textbf{PSNR} $\uparrow$ & \textbf{SAM} $\downarrow$ & \textbf{ERGAS} $\downarrow$ & \textbf{UIQI} $\uparrow$ \\
\hline
CNMF &0.0149&36.55&3.63&1.36&0.9854 \\
HyCoNet & 0.0138&37.17&3.40&0.95&0.9879 \\
MIAE & 0.0109&39.27&\underline{2.66}&1.14&0.9912 \\
EU2ADL   & 0.0115&38.76&3.05&1.06&0.9910\\
FeafusFormer & 0.0122&38.24&3.14&1.12&0.9906\\ 
OTIAS & 0.0112&38.98&2.93&0.96&0.9912 \\
EDIP-Net & \underline{0.0108}&\underline{39.32}&2.81&\underline{0.94}&\underline{0.9922} \\
SCALMU  &\textbf{0.0098}&\textbf{40.15}&\textbf{2.43}&\textbf{0.93}&\textbf{0.9930}\\
\hline
\hline
\end{tabular}
\end{table}

\begin{table}
\centering
\caption{Quantitative evaluation on \textbf{Chikusei} dataset ($\times 8$) comparing CNMF, HyCoNet, MIAE, EU2ADL, FeafusFormer, OTIAS, EDIP-Net and SCALMU (ours). Best/second-best results in \textbf{bold}/\underline{underline}.}
\label{tab_chikusei_x8}
\begin{tabular}{l|ccccc}
\hline
\hline
\textbf{Method} & \textbf{RMSE} $\downarrow$ & \textbf{PSNR} $\uparrow$ & \textbf{SAM} $\downarrow$ & \textbf{ERGAS} $\downarrow$ & \textbf{UIQI} $\uparrow$ \\
\hline
CNMF  & 0.0063&44.05&3.70&0.89&0.9784 \\
HyCoNet  & 0.0059&44.61&3.07&\underline{0.56}&0.9798\\
MIAE     & \underline{0.0041}&\underline{47.79}&\underline{2.75}&0.63&\underline{0.9826} \\
EU2ADL  & 0.0045&46.90&3.13&0.68&0.9814 \\
FeafusFormer & 0.0048&46.35&2.93&0.80&0.9823\\
OTIAS & 0.0051&45.79&3.40&0.82&0.9844 \\
EDIP-Net & 0.0046&46.66&3.27&0.78&0.9837 \\
SCALMU   & \textbf{0.0040}&\textbf{48.03}&\textbf{2.65}&\textbf{0.56}&\textbf{0.9827 }\\
\hline
\hline
\end{tabular}
\end{table}

\begin{table}
\centering
\caption{Quantitative evaluation on \textbf{PRISMA-Paris} dataset ($\times 8$) comparing CNMF, HyCoNet, MIAE, EU2ADL, FeafusFormer, OTIAS, EDIP-Net and SCALMU (ours). Best/second-best results in \textbf{bold}/\underline{underline}.}
\label{tab_prisma_x8}
\begin{tabular}{l|ccccc}
\hline
\hline
\textbf{Method} & \textbf{RMSE} $\downarrow$ & \textbf{PSNR} $\uparrow$ & \textbf{SAM} $\downarrow$ & \textbf{ERGAS} $\downarrow$ & \textbf{UIQI} $\uparrow$ \\
\hline
CNMF  & 0.0123&38.22&5.51&0.97&0.8718 \\
HyCoNet  & 0.0090&40.90&4.69&0.77&0.8785\\
MIAE     & \underline{0.0072}&\underline{42.90}&4.87&0.58&0.8814 \\
EU2ADL  & 0.0073&42.68&\underline{4.82}&\underline{0.57}&\underline{0.8820} \\
FeafusFormer &0.0102&39.86&5.60&0.78&0.8699\\
OTIAS & 0.0080&41.90&5.34&0.62&0.8803 \\
EDIP-Net &  0.0075&42.48&5.16&0.60&0.8818\\
SCALMU          & \textbf{0.0069}&\textbf{43.24}&\textbf{4.42}&\textbf{0.52}&\textbf{0.8822} \\
\hline
\hline
\end{tabular}
\end{table}

\subsection{Evaluation on Real Data}
\label{subsec_real_data}

In this section, we evaluate SCALMU on a real-world dataset and compare it with two high-performing state-of-the-art methods previously considered in the simulated experiments, namely MIAE and EDIP-Net. This experiment aims to assess the practical effectiveness of the different fusion approaches in a realistic scenario, without relying on simulated degradation models. The dataset was acquired by the Ziyuan-1 02D satellite and includes HSI, MSI, and PAN images \cite{li2024real}. In our work, we focus exclusively on the HSI and MSI modalities. The HSI has a native spatial resolution of 30 m (76 bands retained after removing noisy bands), while the MSI offers 10 m resolution across 8 bands. For experiments, we select a $300 \times 300$ urban scene region from the MSI, with its corresponding $100 \times 100$ HSI patch ($\times 3$ scale factor).

Since no high-resolution hyperspectral ground truth is available for real data, the evaluation relies primarily on no-reference quality assessment metrics. In particular, we use the quality with no reference (QNR) index and its associated spectral and spatial distortion components, $D_{\lambda}$ and $D_{S}$, respectively, following \cite{alparone2008multispectral}. For completeness, we also report consistency measures in the HSI and MSI domains through $\mathrm{PSNR}_{HSI}$ and $\mathrm{PSNR}_{MSI}$ which can be computed by reprojecting the fused image into the observed domains using  the degradation operators $\mathbf{P}$ and $\mathbf{R}$ estimated by each method.

\begin{figure*}
    \centering
    \includegraphics[width=0.8\linewidth]{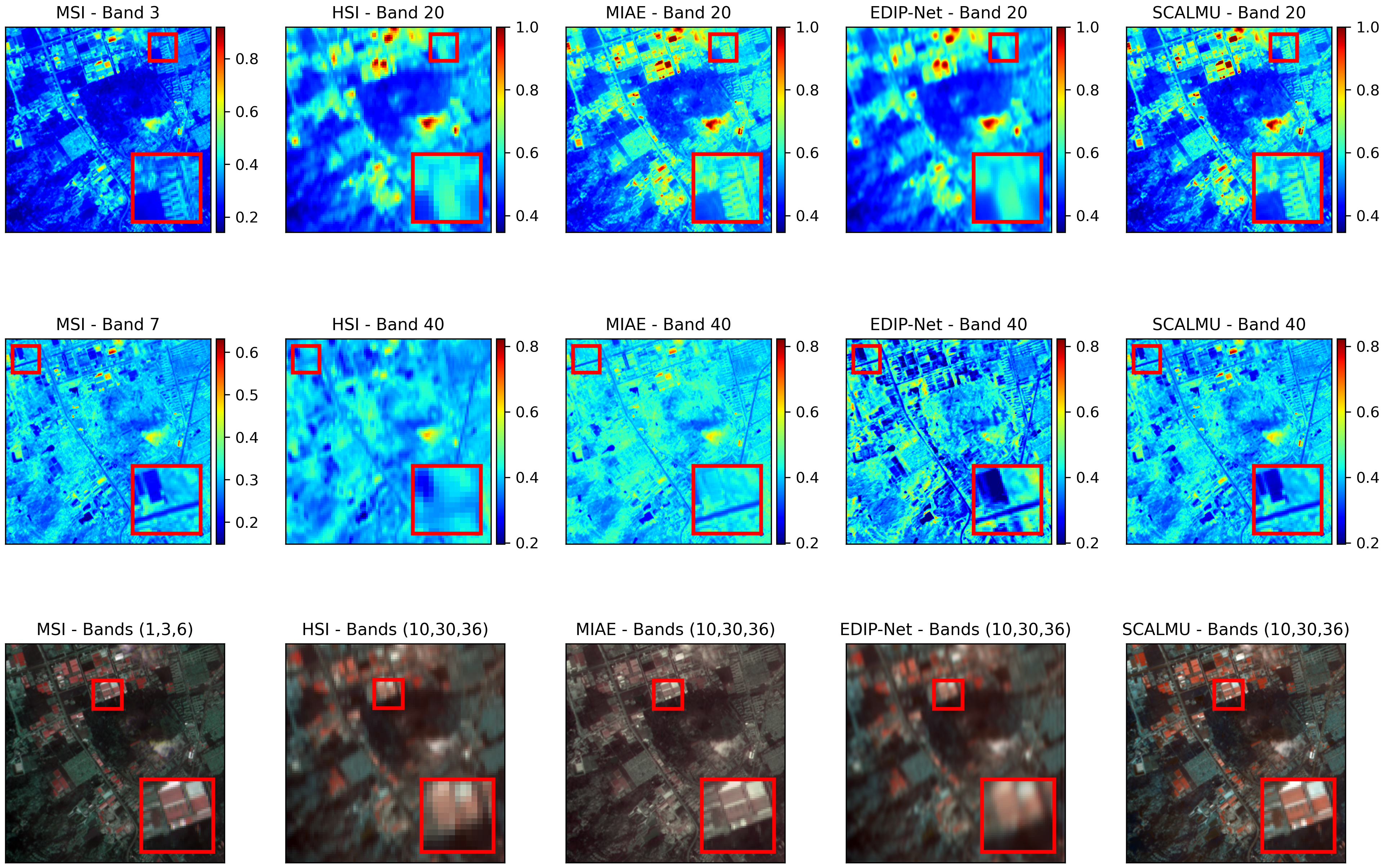}
    \caption{Visual comparison on the real \textbf{Ziyuan-1 02D} dataset ($\times 3$). Column 1 (MSI): bands 3 and 7, and pseudo-color RGB (1, 3, 6). Column 2 (HSI): bands 21 and 40, and pseudo-color RGB (10, 30, 36). Columns 3--5 (SRI): the corresponding fused super-resolved hyperspectral images obtained by MIAE, EDIP-Net, and the proposed SCALMU method.}
    \label{fig_SR_real}
\end{figure*}

\begin{table}
\centering
\setlength{\tabcolsep}{3pt} 
\caption{Quantitative evaluation on \textbf{Ziyuan-1 02D} dataset ($\times 3$) comparing MIAE, EDIP-Net and SCALMU (ours). Best results in \textbf{bold}.}
\label{tab_ziyuan_x3}
\begin{tabular}{l|ccccc}
\hline
\hline
\textbf{Method} & \textbf{$QNR$} $\uparrow$ & \textbf{$D_{\lambda}$} $\downarrow$ & \textbf{$D_S$} $\downarrow$ & \textbf{$PSNR_{HSI}$} $\uparrow$ & \textbf{$PSNR_{MSI}$} $\uparrow$ \\
\hline
MIAE     & 0.9483 & 0.0333& 0.0190& 34.87& 35.33\\
EDIP-Net & 0.9236& \textbf{0.0310}& 0.0468 &34.96 & 32.57\\
SCALMU   & \textbf{0.9578}& 0.0323& \textbf{0.0102}& \textbf{35.00}& \textbf{37.81}\\
\hline
\hline
\end{tabular}
\end{table}

As shown in Table \ref{tab_ziyuan_x3} and Fig. \ref{fig_SR_real}, SCALMU delivers the best overall performance on the real Ziyuan-1 02D dataset, both quantitatively and visually, achieving the best trade-off between spatial reconstruction and spectral preservation. Although EDIP-Net achieves the lowest $D_{\lambda}$, its significantly higher $D_S$ reveals weaker spatial reconstruction, which is also apparent in the visual results. In contrast, MIAE better preserves spatial structures, but at the expense of spectral fidelity. Visually, the super-resolved images produced by SCALMU effectively combine the sharp spatial details from the MSI with the rich spectral information from the HSI, yielding convincing fusion results on this real dataset. Fine urban structures are clearly reconstructed, while spectral signatures remain consistent across the displayed bands and pseudo-color images. Overall, these quantitative and visual comparisons confirm that MIAE tends to emphasize spatial structures at the expense of spectral consistency, whereas EDIP-Net favors spectral fidelity but reconstructs spatial details less effectively. SCALMU provides the best balance between these two aspects, demonstrating its practical relevance for real-world hyperspectral image super-resolution fusion tasks.

\subsection{Ablation Studies}
\label{subsec_Ablation}

We conduct ablation studies on four key aspects in this section: degradation estimation (Section \ref{subsubsec_psf_srf_estimation}), comparing ground-truth versus estimated PSF/SRF; adaptability of the learned updates (Section \ref{subsubsec_adaptability_analysis}), comparing adaptive SCALMU with non-adaptive NALMU; contribution of each adaptive module (Section \ref{subsubsec_analysis_module}), analyzing abundance and spectral adaptive modules for HSI and MSI; and relevance of synthetic dead leaves training data (Section \ref{subsubsec_analysis_DL}), comparing synthetic versus real PRISMA training.

\subsubsection{Degradation estimation}
\label{subsubsec_psf_srf_estimation}

As described in Section \ref{subsec_blindnet}, we designed a degradation estimation network to jointly estimate the PSF and SRF by minimizing the spectral and spatial low-resolution inconsistencies between HSI and MSI. When used upstream of SCALMU, we show that this network produces sufficiently accurate PSF and SRF estimates for our fusion task.

\begin{figure}
    \centering
    \includegraphics[width=1\linewidth]{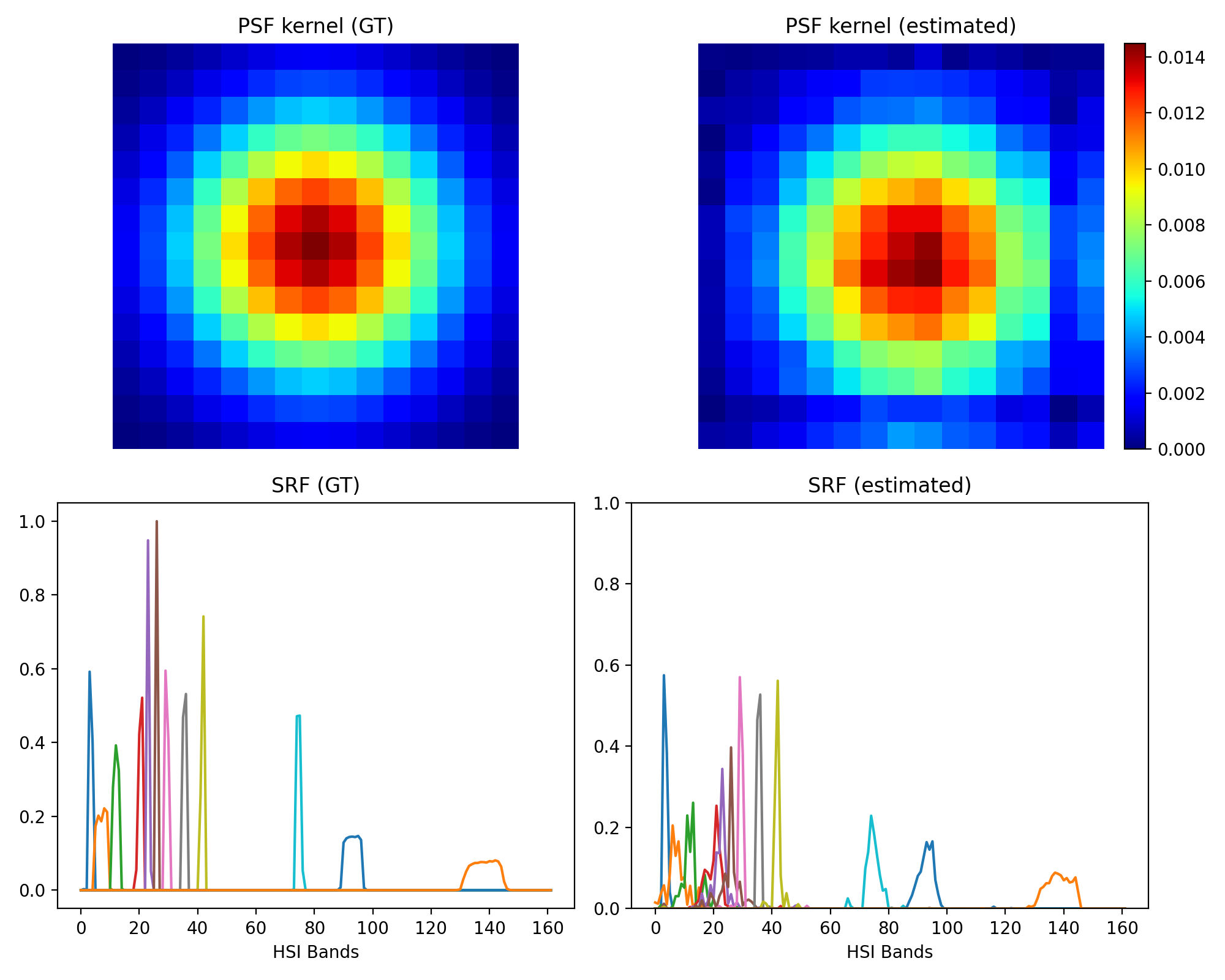}
    \caption{Visual comparison between real and estimated PSF \& SRF on \textbf{Urban} dataset ($\times 8$). The corresponding quantitative estimation errors are PSF RMSE = 0.0011, SRF RMSE = 0.0207, and SRF SAD = 0.3492 rad.}
    \label{fig_P_R_Urban}
\end{figure}

\begin{figure}
    \centering
    \includegraphics[width=1\linewidth]{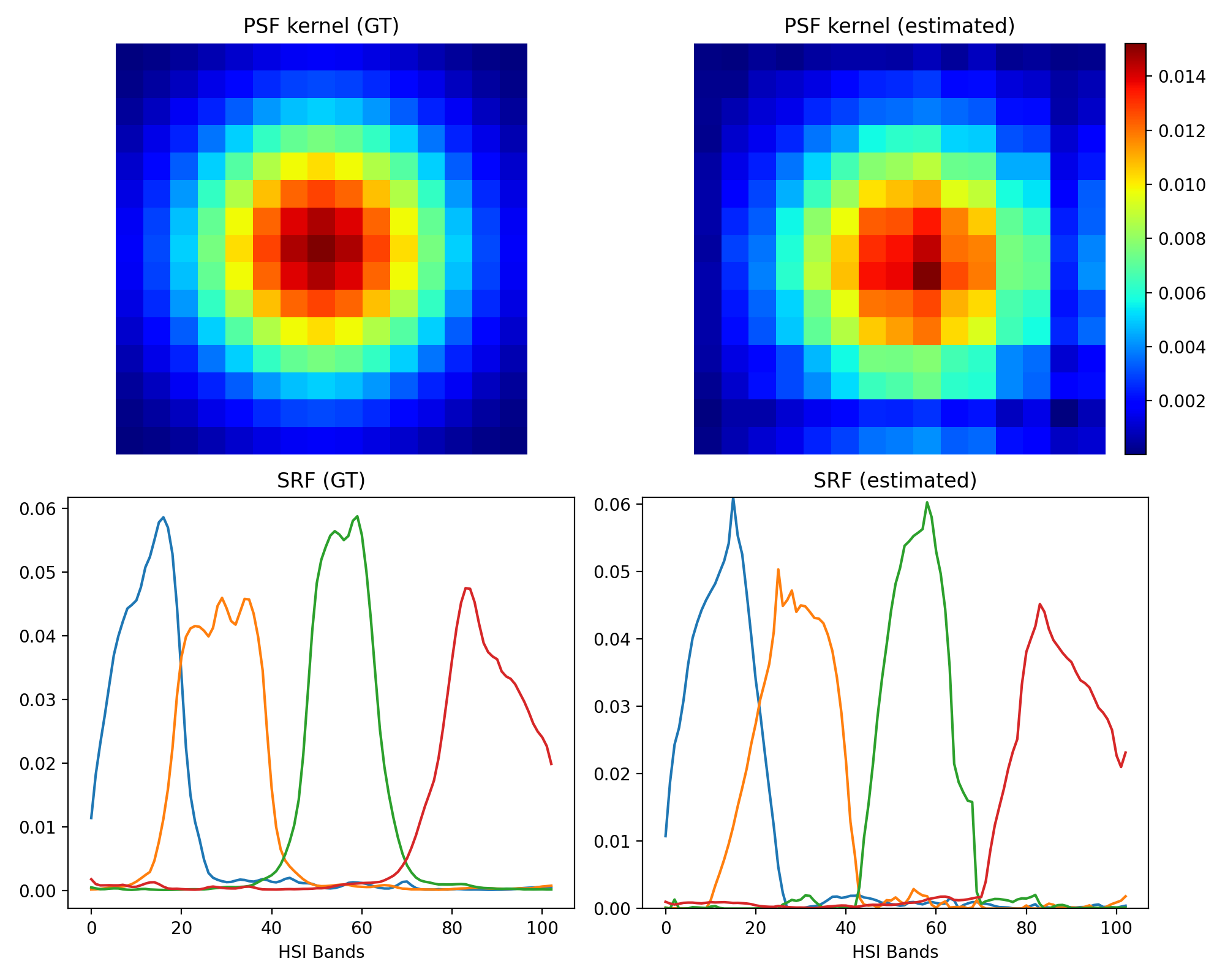}
    \caption{Visual comparison between real and estimated PSF \& SRF on \textbf{PaviaU} dataset ($\times 8$). The corresponding quantitative estimation errors are PSF RMSE = 0.0010, SRF RMSE = 0.0019, and SRF SAD = 0.0965 rad.}
    \label{fig_P_R_PaviaU}
\end{figure}

Figs. \ref{fig_P_R_Urban} and \ref{fig_P_R_PaviaU} illustrate both the visual comparison and the quantitative estimation errors between the ground-truth (GT) and estimated PSF \& SRF on the Urban and PaviaU datasets ($\times 8$), respectively. On the Urban dataset, we further conducted a quantitative ablation study comparing SCALMU using GT degradation parameters (SCALMU non-Blind) versus estimated parameters (SCALMU Blind), as reported in Table \ref{tab_ablation_blind}. The results show nearly identical performances, with only marginal degradation, confirming the effectiveness of our estimator for practical blind fusion scenarios.

\begin{table}
\centering
\setlength{\tabcolsep}{3pt}  
\caption{Quantitative evaluation of SCALMU (Blind) and (non-Blind) on \textbf{Urban} dataset ($\times 8$). Best results in \textbf{bold}.}
\label{tab_ablation_blind}
\begin{tabular}{c|ccccc}
\hline
\hline
\textbf{Method} & \textbf{RMSE} $\downarrow$ & \textbf{PSNR} $\uparrow$ & \textbf{SAM} $\downarrow$ & \textbf{ERGAS} $\downarrow$ & \textbf{UIQI} $\uparrow$ \\
\hline
SCALMU(non-Blind) & \textbf{0.0083}&\textbf{41.52}&\textbf{2.16}&\textbf{0.81}&\textbf{0.9963} \\
SCALMU (Blind) & 0.0084&41.49&2.17&0.82&0.9962 \\
\hline
\hline
\end{tabular}
\end{table}

\subsubsection{Adaptability Analysis}
\label{subsubsec_adaptability_analysis}

\begin{table}
\centering
\setlength{\tabcolsep}{3pt}  
\caption{Ablation study on SCALMU adaptability (Urban dataset, $\times 8$). Best results in \textbf{bold}.}
\label{tab_ablation_Adaptability}
\begin{tabular}{c|ccccc}
\hline
\hline
\textbf{Method} & \textbf{RMSE} $\downarrow$ & \textbf{PSNR} $\uparrow$ & \textbf{SAM} $\downarrow$ & \textbf{ERGAS} $\downarrow$ & \textbf{UIQI} $\uparrow$ \\
\hline
SCALMU (NALMU) &  0.0093&40.61&2.44&0.89&0.9954 \\
SCALMU & \textbf{0.0084}&\textbf{41.49}&\textbf{2.17}&\textbf{0.82}&\textbf{0.9962} \\
\hline
\hline
\end{tabular}
\end{table}
 
Unlike non-adaptive unrolled methods using fixed learned parameters, SCALMU uses lightweight neural networks $\underline{W(\cdot)}$ that dynamically generate the update matrices from input data characteristics, enabling adaptation to unseen spatial-spectral patterns and degradation profiles. Specifically, SCALMU builds upon RALMU \cite{kervazo2026Unrolled} (Recursive Adaptive Learned Multiplicative Updates), which uses input-dependent neural networks for NMF update acceleration. To isolate the benefits of this adaptive design, we perform an ablation by replacing SCALMU's four adaptive neural modules with the fixed matrices of NALMU (Non-Adaptive Learned Multiplicative Updates) \cite{kervazo2024deep}. We consider NALMU here as a representative baseline for classical unrolling, since its unfolded updates rely on fixed learned parameters rather than input-adaptive operators. In NALMU, the four update factors $\underline{W_{A_m}}, \underline{W_{A_h}}, \underline{W_{S_m}}$, and $\underline{W_{S_h}}$ are directly learned as iteration-dependent parameters, with dimensions matching those of $A_m$, $A_h$, $S_m$, and $S_h$, respectively. More precisely, for each unfolding iteration and for each branch, the model learns one fixed abundance-related matrix and one fixed endmember-related matrix, which are stored as trainable parameters and reused at every forward pass. These matrices are initialized with ones, optimized jointly with the other model parameters during training, and constrained to remain positive through the clamping operations used in the multiplicative updates. At test time, they are expanded along the batch dimension and applied identically to all samples. Therefore, unlike SCALMU, NALMU relies on fixed learned update operators that depend only on the training set and not on the specific spatial-spectral content of the input pair. Equivalently, one can view NALMU as learning a set of global multiplicative preconditioners for the unfolded multiplicative updates of a given dataset, without adapting to the spatial structures or degradation characteristics of individual HSI-MSI pairs.

Table \ref{tab_ablation_Adaptability} compares these variants on the Urban dataset ($\times 8$). SCALMU consistently outperforms NALMU across all metrics, confirming that our adaptive design delivers superior train set / test set generalization capacity and reconstruction quality compared to fixed-parameter unrolled approaches.

\subsubsection{Model Analysis}
\label{subsubsec_analysis_module}

\begin{table*}
\centering
\setlength{\tabcolsep}{3pt} 
\caption{Ablation analysis of the proposed SCALMU with different module combinations on \textbf{Urban} dataset ($\times 8$). Best results in \textbf{bold}.}
\label{tab_ablation_W}
\begin{tabular}{c|cccc|ccccc}
\hline
\hline
\textbf{Method} & \underline{\textbf{$W_{A_m}(\cdot)$}} & \underline{\textbf{$W_{S_m}(\cdot)$}} & \underline{\textbf{$W_{A_h}(\cdot)$}} & \underline{\textbf{$W_{S_h}(\cdot)$}} & \textbf{RMSE} $\downarrow$ & \textbf{PSNR} $\uparrow$ & \textbf{SAM} $\downarrow$ & \textbf{ERGAS} $\downarrow$ & \textbf{UIQI} $\uparrow$ \\
\hline
CNMF  & \xmark   & \xmark & \xmark & \xmark & 0.0145&36.79&3.84&1.32&0.9877 \\
SCALMU-A  & \cmark   & \xmark & \cmark & \xmark & 0.0102&39.85&2.79&0.94&0.9939 \\
SCALMU-S  & \xmark & \cmark & \xmark & \cmark & 0.0141&37.02&3.39&1.30&0.9890 \\
SCALMU-MSI  & \cmark & \cmark & \xmark & \xmark & 0.0094&40.58&2.73&0.92&0.9948 \\
SCALMU-HSI  & \xmark & \xmark & \cmark & \cmark & 0.0111&39.11&2.79&0.94&0.9927 \\
SCALMU  & \cmark & \cmark & \cmark & \cmark & \textbf{0.0084}&\textbf{41.49}&\textbf{2.17}&\textbf{0.82}&\textbf{0.9962} \\
\hline
\hline
\end{tabular}
\end{table*}

\begin{figure*}
    \centering
    \includegraphics[width=1\linewidth]{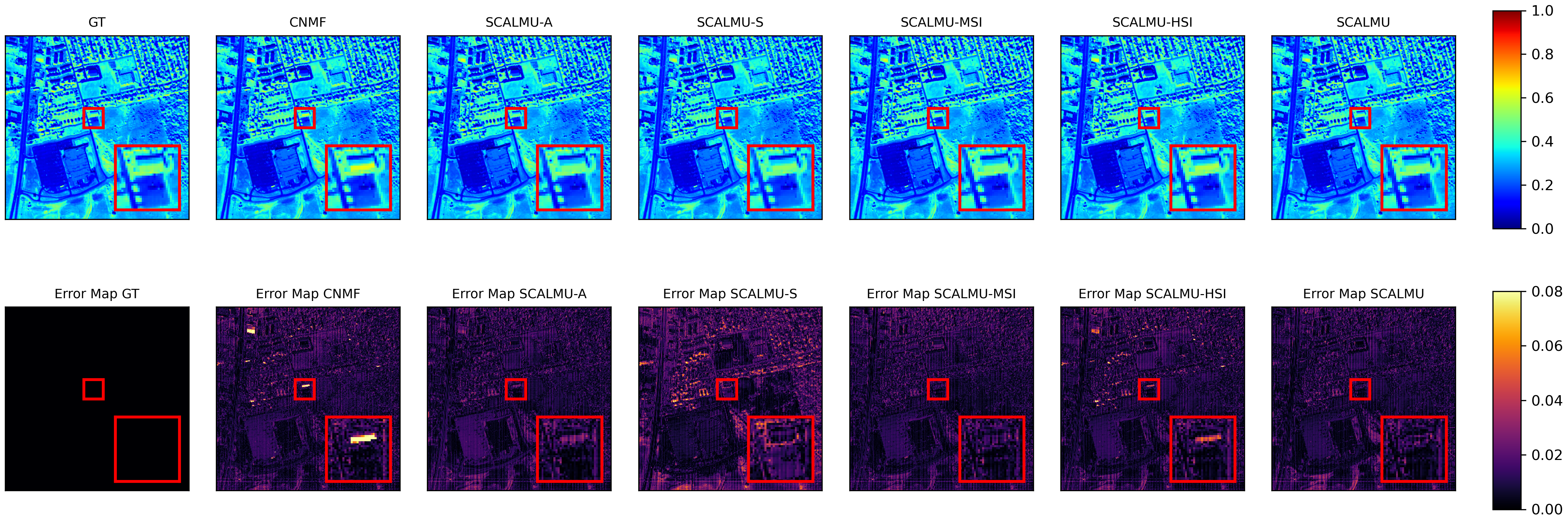}
    \caption{Visual comparison on \textbf{Urban} dataset (band 80). \textbf{Line 1}: reconstructed hyperspectral images (GT, CNMF, SCALMU-A, SCALMU-S, SCALMU-MSI, SCALMU-HSI, SCALMU). \textbf{Line 2}: corresponding absolute error maps $|SR - GT|$. Red rectangles indicate regions zoomed in the insets.} 
    \label{fig_W_studies}
\end{figure*}

To evaluate the contribution of each adaptive module $\underline{W ( \cdot)}$ in SCALMU, we conduct a comprehensive ablation study systematically removing components while keeping the unrolled NMF backbone fixed (Table \ref{tab_ablation_W}, Urban dataset, $\times 8$).

CNMF uses only the classical fixed unrolling without any learned modules, serving as our baseline. SCALMU-A activates abundance-adaptive modules $\underline{W_{A_m}(\cdot)}$ (MSI) and $\underline{W_{A_h}(\cdot)}$ (HSI). SCALMU-S enables spectral-adaptive modules $\underline{W_{S_m}(\cdot)}$ (MSI) and $\underline{W_{S_h}(\cdot)}$ (HSI). SCALMU-MSI combines both MSI modules ($\underline{W_{A_m}(\cdot)}$, $\underline{W_{S_m}(\cdot)}$). SCALMU-HSI uses both HSI modules ($\underline{W_{A_h}(\cdot)}$, $\underline{W_{S_h}(\cdot)}$). Full SCALMU integrates all four modules.

The results in Tab \ref{tab_ablation_W} and Fig \ref{fig_W_studies} reveal distinct contributions. Removing multispectral abundance-adaptive modules $\underline{W_{A_m}(\cdot)}$ causes the most substantial performance degradation, as expected: these modules effectively help extracting high-resolution spatial details from MSI essential for $\times 8$ super-resolution. Spectral modules ($\underline{W_{S_m}(\cdot)}$, $\underline{W_{S_h}(\cdot)}$) provide consistent improvements but have lesser overall impact, primarily enhancing spectral fidelity rather than spatial reconstruction. Notably, as observed in the error maps (Fig \ref{fig_W_studies}), relying solely on spectral modules (SCALMU-S) fails to accurately recover structural information, resulting in pronounced errors particularly around the edges and boundaries of structures. MSI-focused modules yield larger gains than HSI ones, reflecting MSI's critical role in providing high-resolution spatial guidance. Only the full SCALMU configuration achieves optimal synergistic performance across all metrics, validating our modular adaptive design.

\subsubsection{Relevance of synthetic data}
\label{subsubsec_analysis_DL}

\begin{table}
\centering
\setlength{\tabcolsep}{3pt}  
\caption{Quantitative evaluation of SCALMU trained on the synthetic \textbf{dead Leaves} dataset and on the real \textbf{PRISMA} dataset, tested on the \textbf{PRISMA–Paris} scene ($\times 8$). The best results are highlighted in \textbf{bold}.}
\label{tab_ablation_DL}
\begin{tabular}{c|ccccc}
\hline
\hline
\textbf{Training dataset} & \textbf{RMSE} $\downarrow$ & \textbf{PSNR} $\uparrow$ & \textbf{SAM} $\downarrow$ & \textbf{ERGAS} $\downarrow$ & \textbf{UIQI} $\uparrow$ \\
\hline
dead leaves & 0.0069&43.24&4.42&0.52&0.8822 \\
PRISMA & \textbf{0.0067}&\textbf{43.52}&\textbf{4.27}&\textbf{0.51}&\textbf{0.8823} \\
\hline
\hline
\end{tabular}
\end{table}

To validate the effectiveness of our dead leaves synthetic training set, which is generated relying exclusively on the marginal distributions estimated from the PRISMA-Paris scene, we created a real training dataset from PRISMA hyperspectral images over multiple cities (Athens, Copenhagen, Dubai, Rome, etc.), keeping the Paris scene as the test set. We extracted $256 \times 256$ crops across all 230 spectral bands from the other cities, generating 1,000 reference images. Applying the same PSF $\mathbf{P}$ and SRF $\mathbf{R}$ as PRISMA-Paris ensures identical degradation conditions and dataset size ($N=1000$) as our synthetic dead leaves training set, enabling fair comparison.

We trained SCALMU separately on both datasets and evaluated on PRISMA-Paris ($\times 8$), as shown in Table \ref{tab_ablation_DL}. The real PRISMA dataset yields slightly superior performance as expected, but the gap with dead leaves synthetic training remains marginal across all metrics. This demonstrates the remarkable effectiveness of our proposed synthetic data generation process, achieving near-real-data performance while eliminating dependency on scarce ground-truth pairs. It should be noted that this ablation study specifically aims to demonstrate the practical value of synthetic datasets. Indeed, assuming access to a substantial number of real HSI-MSI pairs along with their corresponding high-resolution ground truth is highly unrealistic in real-world remote sensing scenarios. Consequently, the performance achieved with the real PRISMA training dataset should be regarded as a theoretical upper bound rather than a practically attainable baseline.

\subsection{A preliminary study of cross-sensor generalization capacity}
\label{subsec_generalisation}

\begin{table}
\centering
\setlength{\tabcolsep}{1pt}  
\caption{Quantitative evaluation of SCALMU trained on the synthetic \textbf{dead Leaves} dataset and on the real \textbf{PRISMA} dataset, tested on the interpolated \textbf{URBAN} scene ($\times 8$). The best results are highlighted in \textbf{bold}.}
\label{tab_ablation_DL_generalization}
\begin{tabular}{c|c|ccccc}
\hline
\hline
\textbf{Train. dataset} &\textbf{reestim. $\mathbf{PR}$}& \textbf{RMSE} $\downarrow$ & \textbf{PSNR} $\uparrow$ & \textbf{SAM} $\downarrow$ & \textbf{ERGAS} $\downarrow$ & \textbf{UIQI} $\uparrow$ \\
\hline
dead leaves & \xmark &\textbf{0.0118}&\textbf{38.58}&\textbf{2.47}&\textbf{0.98}&\textbf{0.9939} \\
PRISMA & \xmark &0.0139&37.16&3.55&1.15&0.9891 \\
\hline
dead leaves & \cmark & \textbf{0.0112}&\textbf{39.03}&\textbf{2.38}&\textbf{0.96}&\textbf{0.9945}\\
PRISMA & \cmark &0.0132&37.62&3.25&1.09&0.9910 \\
\hline
\hline
\end{tabular}
\end{table}

To further investigate the cross-sensor generalization capacity of synthetic images compared to real ones, we conducted a preliminary experiment  by directly reusing the two SCALMU models trained in Section~\ref{subsubsec_analysis_DL}.  These models, trained respectively on the synthetic dead leaves dataset \emph{generated from PRISMA-Paris abundance marginals} and on the real PRISMA dataset (acquired over multiple cities by PRISMA mission), are tested on the URBAN hyperspectral scene acquired by the HYDICE sensor. HYDICE captures 210 spectral bands within the 400–2500 nm range without spectral overlap, which is comparable to PRISMA’s 230 non-overlapping bands over the same spectral interval. For consistency, the URBAN dataset was linearly interpolated from 210 to 230 bands, degraded using the same PSF $\mathbf{P}$ and SRF $\mathbf{R}$, and used as input for both trained models. The fused outputs were then re-interpolated back to 210 bands, and noisy bands were removed to evaluate the result against the standard 162 bands URBAN reference.

As summarized in Table~\ref{tab_ablation_DL_generalization}, the SCALMU model trained on synthetic dead leaves data outperforms its PRISMA-trained counterpart across all metrics. This holds true regardless of whether the sensor’s PSF and SRF are re-estimated or directly reused from the PRISMA estimation. These preliminary results suggest that while real-data training may lead to a very good fitting to specific sensor characteristics, synthetic supervision exhibits superior cross-sensor generalization. In particular, it appears to improve robustness to differences in spectral responses and modality gaps between heterogeneous hyperspectral and multispectral systems. The role of noise robustness may also be a limiting factor in such transfer settings, but it is not the primary focus of this study and will be investigated in future work. Although further extensive validation is required, these early findings highlight the strong potential of synthetic datasets as a robust, sensor-robust training paradigm.

\subsection{Computational Efficiency}
\label{subsec_Efficiency}

\begin{table}
\centering
\setlength{\tabcolsep}{2pt} 
\caption{Computational cost for different learning-based fusion methods on the Urban dataset (scale $\times 8$).}
\begin{tabular}{l|ccc}
\hline
\hline
\textbf{Method} & \textbf{Dataset Gen.} & \textbf{\#Params} & \textbf{Running time}* \\
\hline
HyCoNet  &  -- & 400K & 45min \\
MIAE  &  --  & 200K & 20min \\
EU2ADL &  -- & 1M & 53min   \\
FeafusFormer & -- & 38M & 1h10 $\mid$ 4s\\
OTIAS &  -- & 3M & 1h50 $\mid$ 4s\\
EDIP-Net &  --  & 12M & 1h40 \\
SCALMU &  11min  & 2M  & 5h40 $\mid$ 3s \\
\hline
\hline
\end{tabular}
\\ 
{\footnotesize *\textit{If 2 running times are given as $t_1 \mid t_2$, $t_1$ corresponds to training time and $t_2$ to inference time. Otherwise, the time corresponds to the processing time needed to fuse one image pair.}}

\label{tab_complexity_comparison}
\end{table}

Table \ref{tab_complexity_comparison} presents the computational costs (dataset generation, parameter count, and running times on V100 GPU) for learning-based fusion methods on the Urban dataset ($\times 8$). The “Dataset Gen.” column indicates the time required to generate the 1,000 synthetic training images used by SCALMU, which are produced using the dead leaves model. Despite longer initial training for its end-to-end synthetic learning, it achieves fast inference per image pair, matching OTIAS while using fewer parameters and delivering better reconstruction quality, whereas single-time autoencoder methods require lengthy retraining per pair that prevents generalization; SCALMU's one-time training thus might enable rapid deployment across datasets with minimal inference overhead and optimal parameter efficiency.

\section{Conclusion}
\label{conclusion}

This paper introduces SCALMU, a novel blind hyperspectral-multispectral fusion framework that unrolls CNMF into adaptive learned layers. Its core CALMU structure augments classical multiplicative updates with four lightweight neural modules that dynamically predict input-dependent learned matrices for MSI/HSI abundances and spectra, preserving CNMF's physical interpretability while enabling data-driven refinement. A lightweight degradation estimation subnetwork jointly predicts PSF and SRF from input pairs during preprocessing. Trained exclusively on dead leaves synthetic data simulating realistic spatial-spectral statistics, SCALMU eliminates paired real-data requirements. It achieves state-of-the-art performance on several hyperspectral datasets at $\times 8$ scale across all hyperspectral quality metrics, surpassing classical and deep learning methods while generalizing robustly to real-world data. Future work will focus on extending SCALMU’s generalization across heterogeneous sensors through the development of a unified fusion framework that remains robust and consistent under diverse sensor characteristics and imaging conditions.

\section*{Acknowledgments}
\noindent The work was partially supported by Agence de l’Innovation de Défense – AID - via Centre Interdisciplinaire d’Etudes pour la Défense et la Sécurité – CIEDS - (project 2023 - ALIA).

\bibliographystyle{IEEEtran}
\bibliography{biblio}

\begin{thebibliography}{10}
\providecommand{\url}[1]{#1}
\csname url@samestyle\endcsname
\providecommand{\newblock}{\relax}
\providecommand{\bibinfo}[2]{#2}
\providecommand{\BIBentrySTDinterwordspacing}{\spaceskip=0pt\relax}
\providecommand{\BIBentryALTinterwordstretchfactor}{4}
\providecommand{\BIBentryALTinterwordspacing}{\spaceskip=\fontdimen2\font plus
\BIBentryALTinterwordstretchfactor\fontdimen3\font minus
  \fontdimen4\font\relax}
\providecommand{\BIBforeignlanguage}[2]{{%
\expandafter\ifx\csname l@#1\endcsname\relax
\typeout{** WARNING: IEEEtran.bst: No hyphenation pattern has been}%
\typeout{** loaded for the language `#1'. Using the pattern for}%
\typeout{** the default language instead.}%
\else
\language=\csname l@#1\endcsname
\fi
#2}}
\providecommand{\BIBdecl}{\relax}
\BIBdecl

\bibitem{fahes2022unrolling}
M.~Fahes, C.~Kervazo, J.~Bobin, and F.~Tupin, ``Unrolling palm for sparse
  semi-blind source separation,'' in \emph{International Conference on Learning
  Representations}, 2022.

\bibitem{wu2022uiu}
X.~Wu, D.~Hong, and J.~Chanussot, ``Uiu-net: U-net in u-net for infrared small
  object detection,'' \emph{IEEE Transactions on Image Processing}, vol.~32,
  pp. 364--376, 2022.

\bibitem{mills2010evaluation}
S.~J. Mills, M.~P.~G. Castro, Z.~Li, J.~Cai, R.~Hayward, L.~Mejias, and R.~A.
  Walker, ``Evaluation of aerial remote sensing techniques for vegetation
  management in power-line corridors,'' \emph{IEEE Transactions on Geoscience
  and Remote Sensing}, vol.~48, no.~9, pp. 3379--3390, 2010.

\bibitem{yao2023extended}
J.~Yao, B.~Zhang, C.~Li, D.~Hong, and J.~Chanussot, ``Extended vision
  transformer (exvit) for land use and land cover classification: A multimodal
  deep learning framework,'' \emph{IEEE Transactions on Geoscience and Remote
  Sensing}, vol.~61, pp. 1--15, 2023.

\bibitem{lu2020recent}
B.~Lu, P.~D. Dao, J.~Liu, Y.~He, and J.~Shang, ``Recent advances of
  hyperspectral imaging technology and applications in agriculture,''
  \emph{Remote Sensing}, vol.~12, no.~16, p. 2659, 2020.

\bibitem{yokoya2017hyperspectral}
N.~Yokoya, C.~Grohnfeldt, and J.~Chanussot, ``Hyperspectral and multispectral
  data fusion: A comparative review of the recent literature,'' \emph{IEEE
  Geoscience and Remote Sensing Magazine}, vol.~5, no.~2, pp. 29--56, 2017.

\bibitem{yokoya2011coupled}
N.~Yokoya, T.~Yairi, and A.~Iwasaki, ``Coupled nonnegative matrix factorization
  unmixing for hyperspectral and multispectral data fusion,'' \emph{IEEE
  Transactions on Geoscience and Remote Sensing}, vol.~50, no.~2, pp. 528--537,
  2011.

\bibitem{zheng2020coupled}
K.~Zheng, L.~Gao, W.~Liao, D.~Hong, B.~Zhang, X.~Cui, and J.~Chanussot,
  ``Coupled convolutional neural network with adaptive response function
  learning for unsupervised hyperspectral super resolution,'' \emph{IEEE
  Transactions on Geoscience and Remote Sensing}, vol.~59, no.~3, pp.
  2487--2502, 2020.

\bibitem{liu2022model}
J.~Liu, Z.~Wu, L.~Xiao, and X.-J. Wu, ``Model inspired autoencoder for
  unsupervised hyperspectral image super-resolution,'' \emph{IEEE Transactions
  on Geoscience and Remote Sensing}, vol.~60, pp. 1--12, 2022.

\bibitem{achddou2023fully}
R.~Achddou, Y.~Gousseau, and S.~Ladjal, ``Fully synthetic training for image
  restoration tasks,'' \emph{Computer Vision and Image Understanding}, vol.
  233, p. 103723, 2023.

\bibitem{xu2026synthetic}
X.~Xu, Y.~Gousseau, C.~Kervazo, and S.~Ladjal, ``Synthetic abundance maps for
  unsupervised super-resolution of hyperspectral remote sensing images,''
  \emph{IEEE Journal of Selected Topics in Applied Earth Observations and
  Remote Sensing}, vol.~19, pp. 13\,507--13\,520, 2026.

\bibitem{zhukov1999unmixing}
B.~Zhukov, D.~Oertel, F.~Lanzl, and G.~Reinhackel, ``Unmixing-based multisensor
  multiresolution image fusion,'' \emph{IEEE Transactions on Geoscience and
  Remote Sensing}, vol.~37, no.~3, pp. 1212--1226, 1999.

\bibitem{eismann2004application}
M.~T. Eismann and R.~C. Hardie, ``Application of the stochastic mixing model to
  hyperspectral resolution enhancement,'' \emph{IEEE Transactions on Geoscience
  and Remote Sensing}, vol.~42, no.~9, pp. 1924--1933, 2004.

\bibitem{hardie2004map}
R.~C. Hardie, M.~T. Eismann, and G.~L. Wilson, ``Map estimation for
  hyperspectral image resolution enhancement using an auxiliary sensor,''
  \emph{IEEE Transactions on Image Processing}, vol.~13, no.~9, pp. 1174--1184,
  2004.

\bibitem{eismann2005hyperspectral}
M.~T. Eismann and R.~C. Hardie, ``Hyperspectral resolution enhancement using
  high-resolution multispectral imagery with arbitrary response functions,''
  \emph{IEEE Transactions on Geoscience and Remote Sensing}, vol.~43, no.~3,
  pp. 455--465, 2005.

\bibitem{kawakami2011high}
R.~Kawakami, Y.~Matsushita, J.~Wright, M.~Ben-Ezra, Y.-W. Tai, and K.~Ikeuchi,
  ``High-resolution hyperspectral imaging via matrix factorization,'' in
  \emph{2011 IEEE/CVF Conference on Computer Vision and Pattern Recognition
  (CVPR)}.\hskip 1em plus 0.5em minus 0.4em\relax IEEE, 2011, pp. 2329--2336.

\bibitem{akhtar2014sparse}
N.~Akhtar, F.~Shafait, and A.~Mian, ``Sparse spatio-spectral representation for
  hyperspectral image super-resolution,'' in \emph{European conference on
  computer vision}.\hskip 1em plus 0.5em minus 0.4em\relax Springer, 2014, pp.
  63--78.

\bibitem{fu2023mixed}
X.~Fu, H.~Liang, and S.~Jia, ``Mixed noise-oriented hyperspectral and
  multispectral image fusion,'' \emph{IEEE Transactions on Geoscience and
  Remote Sensing}, vol.~61, pp. 1--16, 2023.

\bibitem{fu2021fusion}
X.~Fu, S.~Jia, M.~Xu, J.~Zhou, and Q.~Li, ``Fusion of hyperspectral and
  multispectral images accounting for localized inter-image changes,''
  \emph{IEEE Transactions on Geoscience and Remote Sensing}, vol.~60, pp.
  1--18, 2021.

\bibitem{wang2017deep}
C.~Wang, Y.~Liu, X.~Bai, W.~Tang, P.~Lei, and J.~Zhou, ``Deep residual
  convolutional neural network for hyperspectral image super-resolution,'' in
  \emph{International Conference on Image and Graphics}.\hskip 1em plus 0.5em
  minus 0.4em\relax Springer, 2017, pp. 370--380.

\bibitem{xie2019multispectral}
Q.~Xie, M.~Zhou, Q.~Zhao, D.~Meng, W.~Zuo, and Z.~Xu, ``Multispectral and
  hyperspectral image fusion by ms/hs fusion net,'' in \emph{2019 IEEE/CVF
  Conference on Computer Vision and Pattern Recognition (CVPR)}.\hskip 1em plus
  0.5em minus 0.4em\relax IEEE, 2019, pp. 1585--1594.

\bibitem{zhang2020unsupervised}
L.~Zhang, J.~Nie, W.~Wei, Y.~Zhang, S.~Liao, and L.~Shao, ``Unsupervised
  adaptation learning for hyperspectral imagery super-resolution,'' in
  \emph{2020 IEEE/CVF Conference on Computer Vision and Pattern Recognition
  (CVPR)}.\hskip 1em plus 0.5em minus 0.4em\relax IEEE, 2020, pp. 3070--3079.

\bibitem{han2018ssf}
X.-H. Han, B.~Shi, and Y.~Zheng, ``Ssf-cnn: Spatial and spectral fusion with
  cnn for hyperspectral image super-resolution,'' in \emph{2018 25th IEEE
  International Conference on Image Processing (ICIP)}.\hskip 1em plus 0.5em
  minus 0.4em\relax IEEE, 2018, pp. 2506--2510.

\bibitem{dian2020regularizing}
R.~Dian, S.~Li, and X.~Kang, ``Regularizing hyperspectral and multispectral
  image fusion by cnn denoiser,'' \emph{IEEE Transactions on Neural Networks
  and Learning Systems}, vol.~32, no.~3, pp. 1124--1135, 2020.

\bibitem{ran2023guidednet}
R.~Ran, L.-J. Deng, T.-X. Jiang, J.-F. Hu, J.~Chanussot, and G.~Vivone,
  ``Guidednet: A general cnn fusion framework via high-resolution guidance for
  hyperspectral image super-resolution,'' \emph{IEEE Transactions on
  Cybernetics}, vol.~53, no.~7, pp. 4148--4161, 2023.

\bibitem{qu2018unsupervised}
Y.~Qu, H.~Qi, and C.~Kwan, ``Unsupervised sparse dirichlet-net for
  hyperspectral image super-resolution,'' in \emph{2018 IEEE/CVF Conference on
  Computer Vision and Pattern Recognition}.\hskip 1em plus 0.5em minus
  0.4em\relax IEEE, 2018, pp. 2511--2520.

\bibitem{li2022deep}
J.~Li, K.~Zheng, J.~Yao, L.~Gao, and D.~Hong, ``Deep unsupervised blind
  hyperspectral and multispectral data fusion,'' \emph{IEEE Geoscience and
  Remote Sensing Letters}, vol.~19, pp. 1--5, 2022.

\bibitem{li2023busifusion}
J.~Li, Y.~Li, C.~Wang, X.~Ye, and W.~Heidrich, ``Busifusion: Blind unsupervised
  single image fusion of hyperspectral and rgb images,'' \emph{IEEE
  Transactions on Computational Imaging}, vol.~9, pp. 94--105, 2023.

\bibitem{gao2023enhanced}
L.~Gao, J.~Li, K.~Zheng, and X.~Jia, ``Enhanced autoencoders with
  attention-embedded degradation learning for unsupervised hyperspectral image
  super-resolution,'' \emph{IEEE Transactions on Geoscience and Remote
  Sensing}, vol.~61, pp. 1--17, 2023.

\bibitem{li2023hyperspectral}
S.~Li, Y.~Tian, C.~Wang, H.~Wu, and S.~Zheng, ``Hyperspectral image
  super-resolution network based on cross-scale nonlocal attention,''
  \emph{IEEE Transactions on Geoscience and Remote Sensing}, vol.~61, pp.
  1--15, 2023.

\bibitem{hong2023decoupled}
D.~Hong, J.~Yao, C.~Li, D.~Meng, N.~Yokoya, and J.~Chanussot,
  ``Decoupled-and-coupled networks: Self-supervised hyperspectral image
  super-resolution with subpixel fusion,'' \emph{IEEE Transactions on
  Geoscience and Remote Sensing}, vol.~61, pp. 1--12, 2023.

\bibitem{zhang2024unsupervised}
L.~Zhang, J.~Nie, W.~Wei, and Y.~Zhang, ``Unsupervised test-time adaptation
  learning for effective hyperspectral image super-resolution with unknown
  degeneration,'' \emph{IEEE Transactions on Pattern Analysis and Machine
  Intelligence}, vol.~46, no.~7, pp. 5008--5025, 2024.

\bibitem{li2022symmetrical}
Q.~Li, M.~Gong, Y.~Yuan, and Q.~Wang, ``Symmetrical feature propagation network
  for hyperspectral image super-resolution,'' \emph{IEEE Transactions on
  Geoscience and Remote Sensing}, vol.~60, pp. 1--12, 2022.

\bibitem{hu2022fusformer}
J.-F. Hu, T.-Z. Huang, L.-J. Deng, H.-X. Dou, D.~Hong, and G.~Vivone,
  ``Fusformer: A transformer-based fusion network for hyperspectral image
  super-resolution,'' \emph{IEEE Geoscience and Remote Sensing Letters},
  vol.~19, pp. 1--5, 2022.

\bibitem{cao2024unsupervised}
X.~Cao, Y.~Lian, K.~Wang, C.~Ma, and X.~Xu, ``Unsupervised hybrid network of
  transformer and cnn for blind hyperspectral and multispectral image fusion,''
  \emph{IEEE Transactions on Geoscience and Remote Sensing}, vol.~62, pp.
  1--15, 2024.

\bibitem{wu2023hsr}
C.~Wu, D.~Wang, Y.~Bai, H.~Mao, Y.~Li, and Q.~Shen, ``Hsr-diff: Hyperspectral
  image super-resolution via conditional diffusion models,'' in \emph{2023
  IEEE/CVF International Conference on Computer Vision (ICCV)}.\hskip 1em plus
  0.5em minus 0.4em\relax IEEE, 2023, pp. 7060--7070.

\bibitem{li2025enhanced}
J.~Li, K.~Zheng, L.~Gao, Z.~Han, Z.~Li, and J.~Chanussot, ``Enhanced deep image
  prior for unsupervised hyperspectral image super-resolution,'' \emph{IEEE
  Transactions on Geoscience and Remote Sensing}, 2025.

\bibitem{deng2025otias}
S.~Deng, J.~Ma, L.-J. Deng, and P.~Wei, ``Otias: Octree implicit adaptive
  sampling for multispectral and hyperspectral image fusion,'' in
  \emph{Proceedings of the AAAI Conference on Artificial Intelligence},
  vol.~39, no.~3, 2025, pp. 2708--2716.

\bibitem{dosovitskiy2015flownet}
A.~Dosovitskiy, P.~Fischer, E.~Ilg, P.~H{\"a}usser, C.~Hazirbas, V.~Golkov,
  P.~van~der Smagt, D.~Cremers, and T.~Brox, ``Flownet: Learning optical flow
  with convolutional networks,'' in \emph{2015 IEEE International Conference on
  Computer Vision (ICCV)}.\hskip 1em plus 0.5em minus 0.4em\relax IEEE, 2015,
  pp. 2758--2766.

\bibitem{tian2023stablerep}
Y.~Tian, L.~Fan, P.~Isola, H.~Chang, and D.~Krishnan, ``Stablerep: Synthetic
  images from text-to-image models make strong visual representation
  learners,'' \emph{Advances in Neural Information Processing Systems},
  vol.~36, pp. 48\,382--48\,402, 2023.

\bibitem{cross1983markov}
G.~R. Cross and A.~K. Jain, ``Markov random field texture models,'' \emph{IEEE
  Transactions on Pattern Analysis and Machine Intelligence}, no.~1, pp.
  25--39, 1983.

\bibitem{heeger1995pyramid}
D.~J. Heeger and J.~R. Bergen, ``Pyramid-based texture analysis/synthesis,'' in
  \emph{22nd International ACM Conference on Computer Graphics and Interactive
  Techniques}, 1995, pp. 229--238.

\bibitem{galerne2011micro}
B.~Galerne, Y.~Gousseau, and J.-M. Morel, ``Micro-texture synthesis by phase
  randomization,'' \emph{Image Processing On Line}, vol.~1, pp. 213--237, 2011.

\bibitem{matheron1968modele}
G.~Matheron, ``Modele s{\'e}quentiel de partition al{\'e}atoire,'' Technical
  report, CMM, Tech. Rep., 1968.

\bibitem{alvarez1999size}
L.~Alvarez, Y.~Gousseau, and J.-M. Morel, ``The size of objects in natural and
  artificial images,'' in \emph{Advances in Imaging and Electron
  Physics}.\hskip 1em plus 0.5em minus 0.4em\relax Elsevier, 1999, vol. 111,
  pp. 167--242.

\bibitem{gousseau2003dead}
Y.~Gousseau and F.~Roueff, ``The dead leaves model: general results and limits
  at small scales,'' \emph{arXiv preprint math/0312035}, 2003.

\bibitem{bordenave2006dead}
C.~Bordenave, Y.~Gousseau, and F.~Roueff, ``The dead leaves model: a general
  tessellation modeling occlusion,'' \emph{Advances in Applied Probability},
  vol.~38, no.~1, pp. 31--46, 2006.

\bibitem{lee1999learning}
D.~D. Lee and H.~S. Seung, ``Learning the parts of objects by non-negative
  matrix factorization,'' \emph{Nature}, vol. 401, no. 6755, pp. 788--791,
  1999.

\bibitem{kervazo2026Unrolled}
C.~Kervazo and J.~E. Cohen, ``Unrolled multiplicative updates for nonnegative
  matrix factorization applied to hyperspectral unmixing,'' 2026, working paper
  or preprint.

\bibitem{lee2001occlusion}
A.~B. Lee, D.~Mumford, and J.~Huang, ``Occlusion models for natural images: A
  statistical study of a scale-invariant dead leaves model,''
  \emph{International Journal of Computer Vision}, vol.~41, pp. 35--59, 2001.

\bibitem{gousseau2007modeling}
Y.~Gousseau and F.~Roueff, ``Modeling occlusion and scaling in natural
  images,'' \emph{Multiscale Modeling \& Simulation}, vol.~6, no.~1, pp.
  105--134, 2007.

\bibitem{kendall1999perfect}
W.~S. Kendall and E.~Th{\"o}nnes, ``Perfect simulation in stochastic
  geometry,'' \emph{Pattern Recognition}, vol.~32, no.~9, pp. 1569--1586, 1999.

\bibitem{cogliati2021prisma}
S.~Cogliati, F.~Sarti, L.~Chiarantini, M.~Cosi, R.~Lorusso, E.~Lopinto,
  F.~Miglietta, L.~Genesio, L.~Guanter, A.~Damm \emph{et~al.}, ``The prisma
  imaging spectroscopy mission: overview and first performance analysis,''
  \emph{Remote Sensing of Environment}, vol. 262, p. 112499, 2021.

\bibitem{wald1997fusion}
L.~Wald, T.~Ranchin, and M.~Mangolini, ``Fusion of satellite images of
  different spatial resolutions: Assessing the quality of resulting images,''
  \emph{Photogrammetric Engineering and Remote Sensing}, vol.~63, no.~6, pp.
  691--699, 1997.

\bibitem{li2024real}
S.~Li, ``Real hsi-msi-pan image dataset for the
  hyperspectral/multi-spectral/panchromatic image fusion and super-resolution
  fields,'' \emph{arXiv preprint arXiv:2407.02387}, 2024.

\bibitem{alparone2008multispectral}
L.~Alparone, B.~Aiazzi, S.~Baronti, A.~Garzelli, F.~Nencini, and M.~Selva,
  ``Multispectral and panchromatic data fusion assessment without reference,''
  \emph{Photogrammetric Engineering \& Remote Sensing}, vol.~74, no.~2, pp.
  193--200, 2008.

\bibitem{kervazo2024deep}
C.~Kervazo, A.~Chetoui, and J.~E. Cohen, ``Deep unrolling of the multiplicative
  updates algorithm for blind source separation, with application to
  hyperspectral unmixing,'' in \emph{2024 32nd European Signal Processing
  Conference (EUSIPCO)}.\hskip 1em plus 0.5em minus 0.4em\relax IEEE, 2024, pp.
  656--660.

\end{thebibliography}


\begin{IEEEbiography}[{\includegraphics[width=1in,height=1.25in,clip,keepaspectratio]{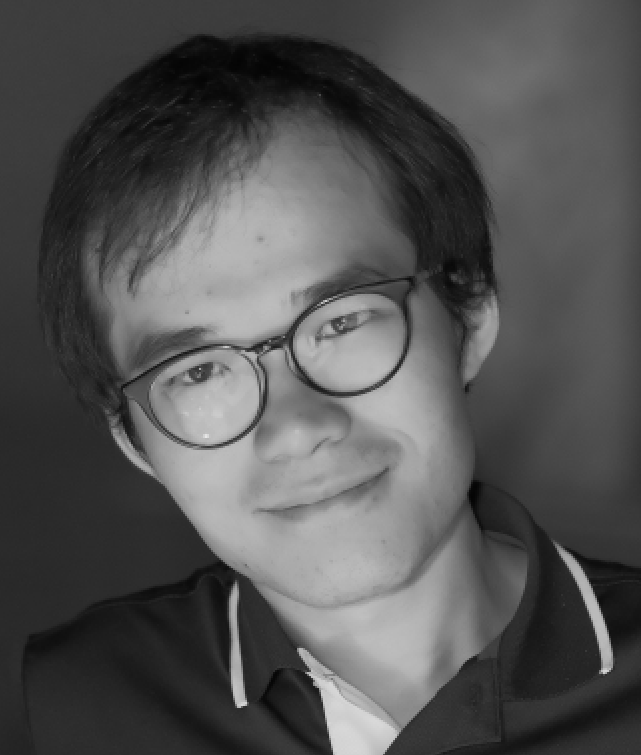}}]{Xinxin Xu} received the engineering degree from Institut d'Optique Graduate School, France, in 2023. He is currently pursuing the Ph.D. degree with LTCI, Télécom Paris, Institut Polytechnique de Paris, Palaiseau, France, under the supervision of Yann Gousseau, Christophe Kervazo and Saïd Ladjal. 

His research focuses on hyperspectral image super-resolution, with applications in remote sensing.\end{IEEEbiography}

\begin{IEEEbiography}[{\includegraphics[width=1in,height=1.25in,clip,keepaspectratio]{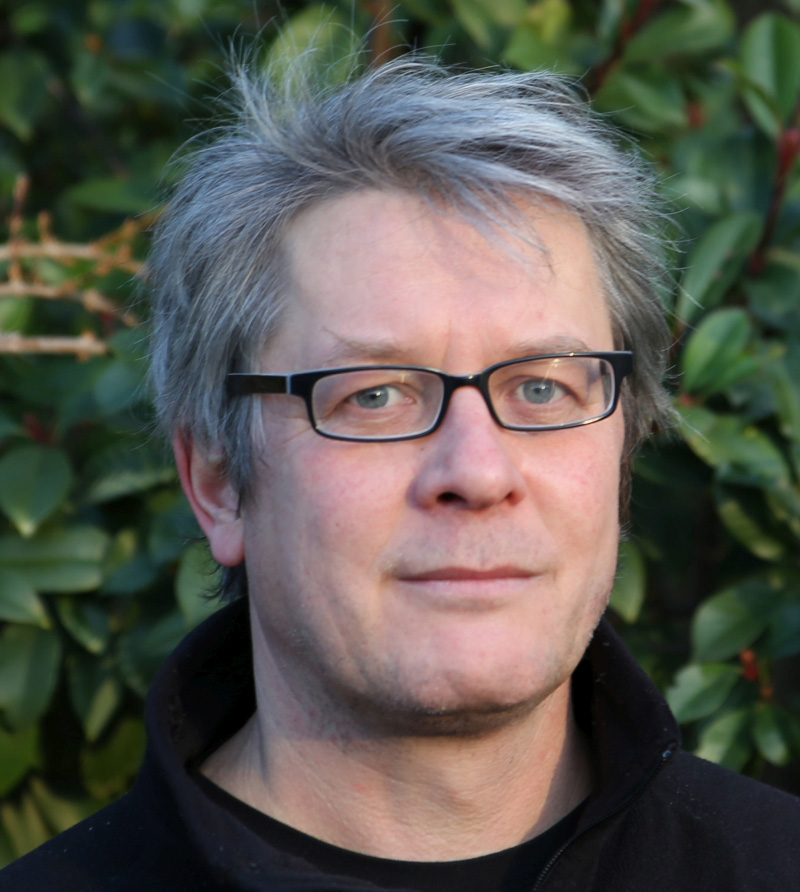}}]{Yann Gousseau} received the B.Eng. degree from the École Centrale de Paris, Châtenay-Malabry, France, the Part III of the Mathematical Tripos degree from the University of Cambridge, Cambridge, U.K., in 1995, and the Ph.D. degree in applied mathematics from the University of Paris-Dauphine, Paris, France, in 2000.

He is currently a Professor with Télécom Paris, Palaiseau, France. His research interests include the mathematical modeling of natural images and textures, generative models, computer vision, image, and video processing.\end{IEEEbiography}

\begin{IEEEbiography}
[{\includegraphics[width=1in,height=1.25in,clip,keepaspectratio]{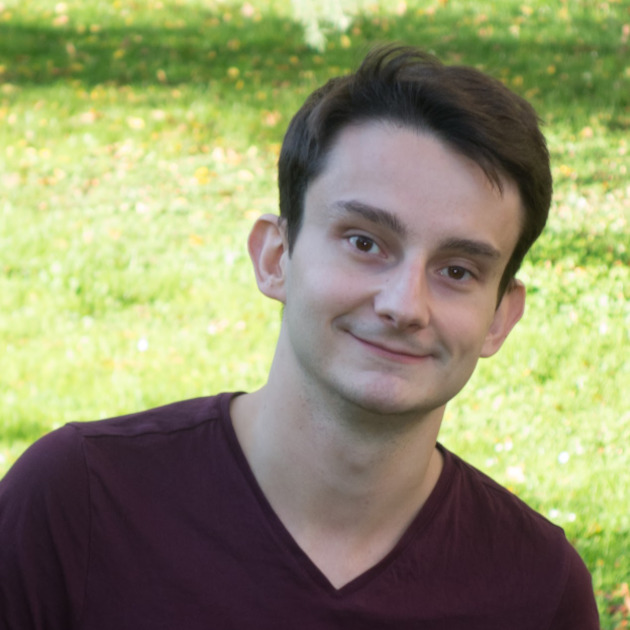}}]{Christophe Kervazo}
 received the engineering degree from Supélec, France, in 2015, as well as the Master of Sciences degree from Georgia Institute of Technology, Atlanta, USA. He did is PhD at CEA Saclay, Gif-sur-Yvette, France , in 2019.
 
 He is currently associate professor at Télécom Paris, Palaiseau, France, where he works on imaging applications. His research focuses on interpretable deep learning, both from the neural networks architecture point of view and the reliability of their outputs. His applications include remote sensing and medical imaging. \end{IEEEbiography}

\begin{IEEEbiography}
[{\includegraphics[width=1in,height=1.25in,clip,keepaspectratio]{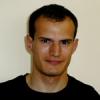}}]{Saïd Ladjal}
 received a diplôme de magister from École normale supérieure in 2000 including a Masters degree in computer science, engineering degree from Télécom Paris in 2002 and Ph.D degree in applied mathematics from École normale supérieure de Cachan in 2005.

 He is currently a Professor with Télécom Paris. His research interests are on mathematical modeling for images and computational photography with applications to remote sensing, general image restoration and medical imaging.  \end{IEEEbiography} 

\end{document}